\documentclass[graybox]{svmult}

\usepackage{caption}
\usepackage{type1cm}        % activate if the above 3 fonts are
\usepackage{makeidx}         % allows index generation
\usepackage{graphicx}        % standard LaTeX graphics tool
\usepackage{multicol}        % used for the two-column index
\usepackage[bottom]{footmisc}% places footnotes at page bottom

\usepackage{newtxtext}       % 
\usepackage{newtxmath}       % selects Times Roman as basic font

\usepackage{ragged2e}
\usepackage{subfigure}

\usepackage{pdflscape}

\makeindex             % used for the subject index
\begin{document}

\title*{Software-Defined Multi-domain Tactical Networks: Foundations and Future
Directions}
% Use \titlerunning{Short Title} for an abbreviated version of
% your contribution title if the original one is too long
\author{Redowan Mahmud, Adel N. Toosi, Maria Alejandra Rodriguez, Sharat Chandra Madanapalli, Vijay Sivaraman, Len Sciacca, Christos Sioutis and Rajkumar Buyya}
% Use \authorrunning{Short Title} for an abbreviated version of
% your contribution title if the original one is too long
\institute{
Redowan Mahmud, Maria Alejandra Rodriguez, Len Sciacca, Rajkumar Buyya \at The University of Melbourne, Australia. 
\and 
Adel N. Toosi \at Monash University, Australia.
\and  
Sharat Chandra Madanapalli, Vijay Sivaraman \at The University of New South Wales, Australia.
\and 
Christos Sioutis \at Defence Science and Technology, Department of Defence,Australia.}
%
% Use the package "url.sty" to avoid
% problems with special characters
% used in your e-mail or web address
\renewcommand\rightmark{Mahmud et al.}
\renewcommand\leftmark{Software-Defined Multi-domain Tactical Networks}
\maketitle
\abstract{Software Defined Networking (SDN) has emerged as a programmable approach for provisioning and managing network resources by defining a clear separation between the control and data forwarding planes. Nowadays SDN has gained significant attention in the military domain. Its use in the battlefield communication facilitates the end-to-end interactions and assists the exploitation of edge computing resources for processing data in the proximity. However, there are still various challenges related to the security and interoperability among several heterogeneous, dynamic, intermittent, and data packet technologies like multi-bearer network (MBN) that need to be addressed to leverage the benefits of SDN in tactical environments. In this chapter, we explicitly analyse these challenges and review the current research initiatives in SDN-enabled tactical networks. We also present a taxonomy on SDN-based tactical network orchestration according to the identified challenges and map the existing works to the taxonomy aiming at determining the research gaps and suggesting future directions.}

\section{Introduction}
\label{sec:1}
Networking and communication technologies, especially for competitive and resource constrained environments like battlefields, are continuously evolving \cite{TradeOff}. Similarly, the sensitivity to latency varies significantly between different military applications. For example, the data packet delivery deadline for an application assisting unmanned aerial vehicle (UAV) navigation is quite stringent compared to that of a slow speed on-ground military vehicle. On the other hand, the lifetime and the amount of data handled by a sense-process-actuate cycle-based application is quite shorter than an application broadcasting wartime video stream \cite{EdgeAffinity}. Moreover, military applications require a variety of networking support such as narrowband, broadband, and mobile services to operate. For example, the applications serving tactical wallet Radio Frequency Identification (RFID) and military vehicle Remote Keyless Entry (RKE) harness narrowband services to meet their instantaneous demands. Conversely, the application sharing satellite images need broadband services for higher transmission capacity. If the underlying network is unable to satisfy such diverse requirements of a military application, its QoS (e.g., throughput, response time, and packet loss rate) is expected to degrade and the consequences of QoS degradation for any military application can be devastating during military operations \cite{81}. Therefore, the satisfaction of QoS for military applications is crucial in tactical environments. It also urges the network infrastructure to be adaptive so that any change in the application's QoS requirements can be handled \cite{Latency}. 
\begin{figure}[!t]
\centering 
\captionsetup{justification=centering}
\includegraphics[width=\textwidth]{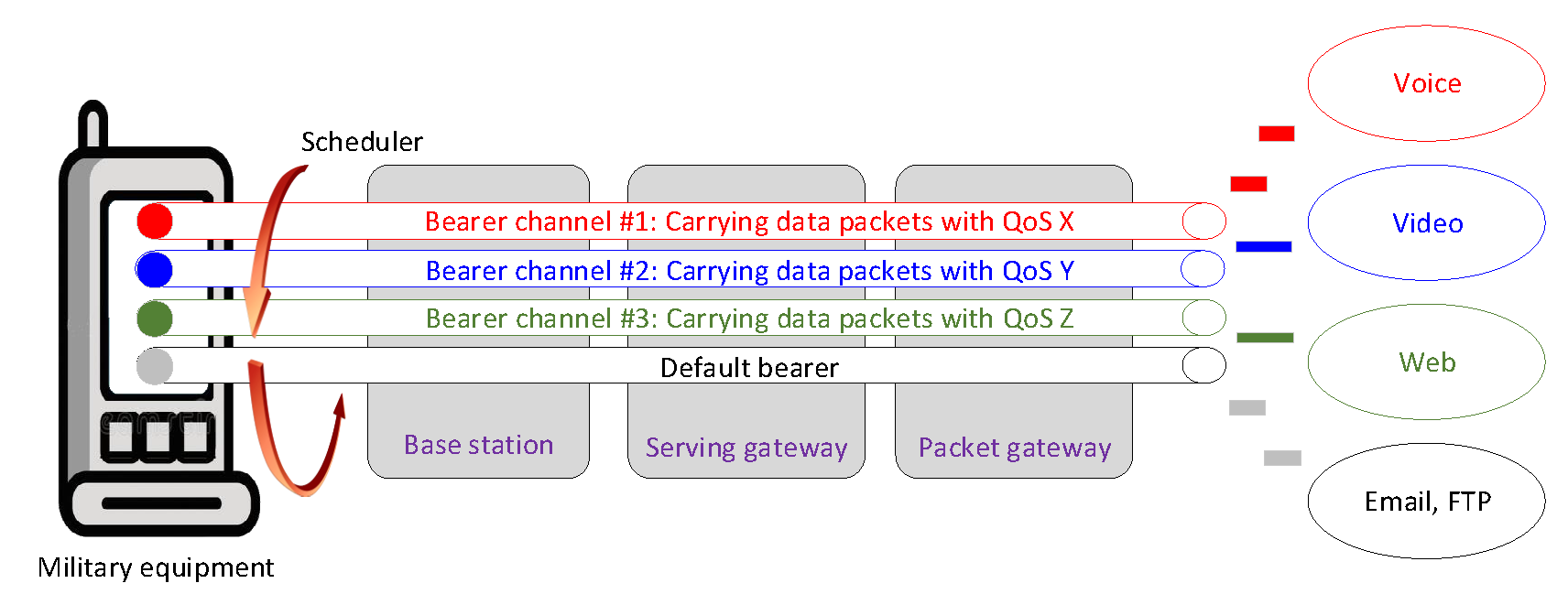}
\caption{Multi-bearer network with differentiated services}
\label{Fig:diffServ}
\end{figure}
\par
\begin{figure}[!t]
\centering 
\captionsetup{justification=centering}
\includegraphics[width=\textwidth]{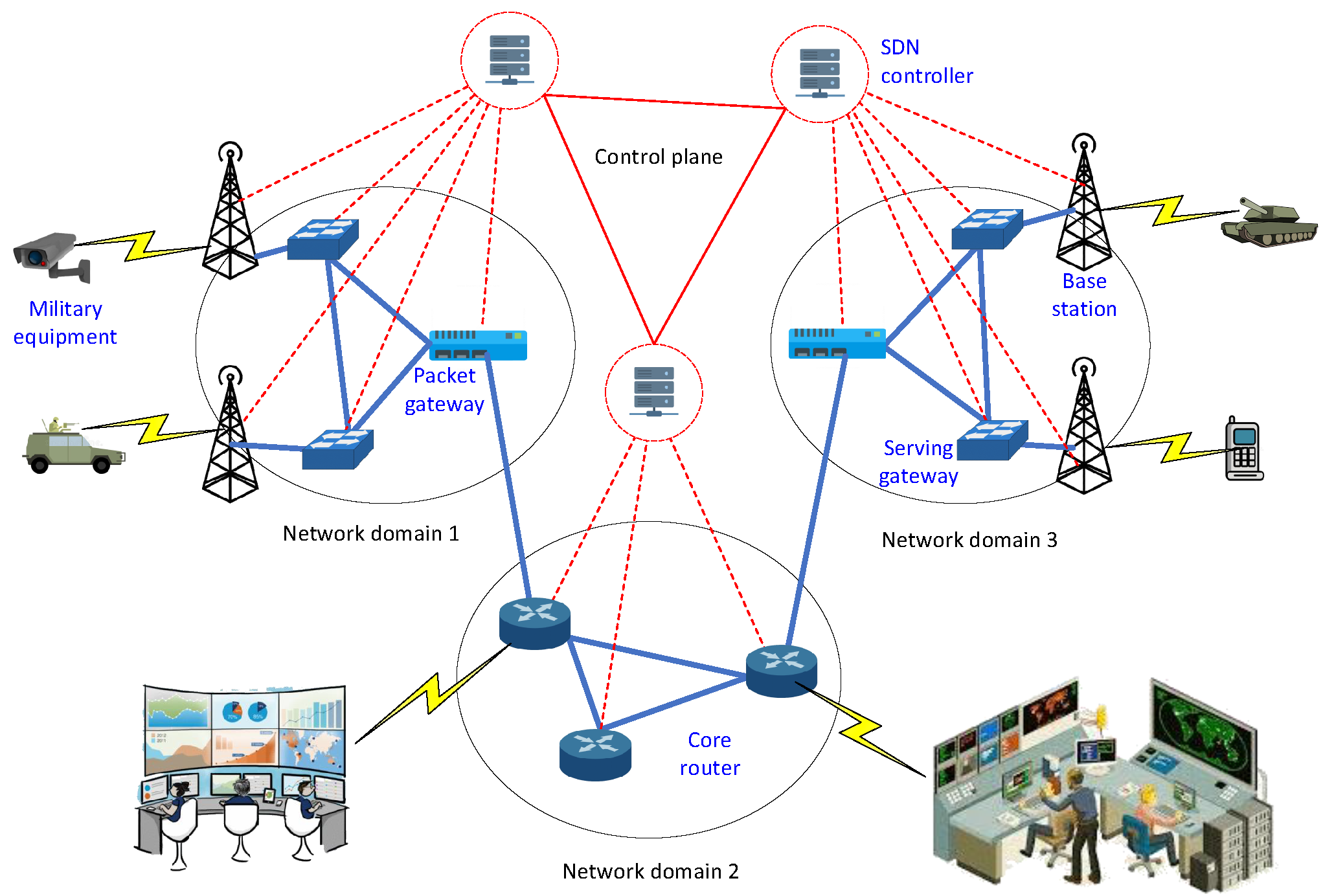}
\caption{A prospective SDN-enabled multi bearer network}
\label{Fig.MBNSDN}
\end{figure}

 Existing data packet technologies, for example multi-bearer network (MBN)can address these requirements to some extent \cite{14}. MBN possesses the capability of carrying data packets via alternative bearer channels as per their QoS requirements. It is complemented by Differentiated Services (DiffServ) that classifies and manages different types of IP traffic (e.g. voice, video, text) flowing over a given network (Fig. \ref{Fig:diffServ}). Nevertheless, communication among multiple nodes within and beyond the battlefields are no longer simply point-to-point. It can be point-to-multipoint and multipoint-to-multipoint as well. In such cases, the realization of MBN incurs additional operational expenses. Moreover, the lack of fair distribution of network resources among the bearer channels can result in severe resource underutilization which is unacceptable for both network operators and military application users \cite{67}. Additionally, the sole advancement of the underlying network is not sufficient to ensure robustness within the multi-domain military operations. It also requires systematic and unified coordination with the computing systems such as fog, mobile edge and cloud infrastructure \cite{Profit}. Therefore, to address these issues and limitations, it is preferable to extend the concept of SDN in tactical networks. Fig.\ref{Fig.MBNSDN} depicts a prospective structure of SDN in military communications.

SDN promotes dynamic provisioning and reconfiguration of network resources by separating the control plane from the data plane \cite{118}. The control plane consists of a logically centralized entity called the SDN controller, which has a global view of the network and makes decisions about how the data packets should flow through the network. Conversely, the data plane consists of network nodes such as routers/switches that actually move packets from one place to another. SDN facilitates virtualization on top of the physical network so that users can implement end-to-end overlays and segment the network traffic. Such logical partitioning also assists the service providers and network operators to provision a separate virtual network with specific policies which consequently complements the objective of MBN and edge computation. 

\subsection{Research Questions and Challenges}
In the context of battlefields and tactical applications, the integration of SDN and MBN is subjected to various heterogeneous, intermittent, and ad-hoc communications with diverse traffic patterns and security requirements. These inherent constraints trigger the following research questions that should be addressed to exploit the combined benefits of SDN and MBN. 
\begin{enumerate}
\item \textit{How can SDN-based solutions be extended to MBN, including wireless networks?}
\begin{itemize}
\item[] Most of the existing SDN-based solutions are applicable to wired networks \cite{EAI}. On the other hand, SDN operations in wireless networks is complicated due to the presence of a large number of unsettled access points. There is also a high possibility of data packet collisions sent by the mutually out-of-range access points. Moreover, the dependency on centralized network controllers is not feasible for latency-sensitive military applications and can expose the whole system to single point of failure problem. 
\end{itemize}
\item \textit{How can SDN be employed for securely and dynamically managing traffic of multiple security classifications, to handle traffic of different sensitivities and access policies, in an environment that includes legacy applications?}
\begin{itemize}
\item[] There are 5 types of classified information including official, protected, secret and top secret that can be transferred during any military communications \cite{MDPI}. However, the security class of information can change dynamically according to the context of the physical environment. For example, the mobilization plan of a fleet can turn from protected to top secret during wartimes. To handle the traffic of such classified information with compatible security features and access flexibility, a consistent inspection of the data packets and environmental context is required. Nevertheless, this approach can expose sensitive traffic data to various untrusted SDN controllers. On the other hand, there exist numerous legacy military applications that still follow the traditional monolithic architecture and provide limited scope to implement SDN-based approaches and resist the secured traffic management and packet inspection.
\end{itemize}   
\item \textit{How can time-sensitive traffic be managed by a multi-bearer SDN, particularly when the time-sensitive channels are required on demand for only limited windows in time?}
\begin{itemize}
\item[] Sensitivity to latency varies between military applications. In such cases, the proactive quantification of QoS requirements and their efficient allocation to the network resources without making over and under provisioning problems are very important \cite{Context}. However, due to less reaction time and variations of resource demands, such SDN-assisted support is difficult to ensure in the battlefield communications.  
\end{itemize} 
\item \textit{How might distributed applications be enhanced with network awareness and control, potentially through coupling to SDN, to make warfighting functions more resilient to degraded network conditions or increased demand on limited network resources?}
\begin{itemize}
\item[] Unlike single-process applications, the components of distributed applications run on multiple hosts simultaneously and process a given task in a collaborative manner \cite{newSurvey}. This consequently helps in attaining scalability and fault tolerance. However, the physical distribution of the components makes the use of networking resources essential in enabling communication and coordination between components. This communication overhead can greatly hinder real-time, latency-sensitive interactions \cite{Realtime}. The ability to have fine-grained control that facilitates the dynamic reconfiguration of network resources to suit distributed applications' needs can greatly improve their resilience and performance. The distributed management of applications is also complex as it requires a fine-grained control over the execution of application components deployed in heterogeneous computing and networking domains \cite{FGCS}.  
\end{itemize}
\begin{figure}[!t]
\centering 
\captionsetup{justification=centering}
\includegraphics[width=\textwidth]{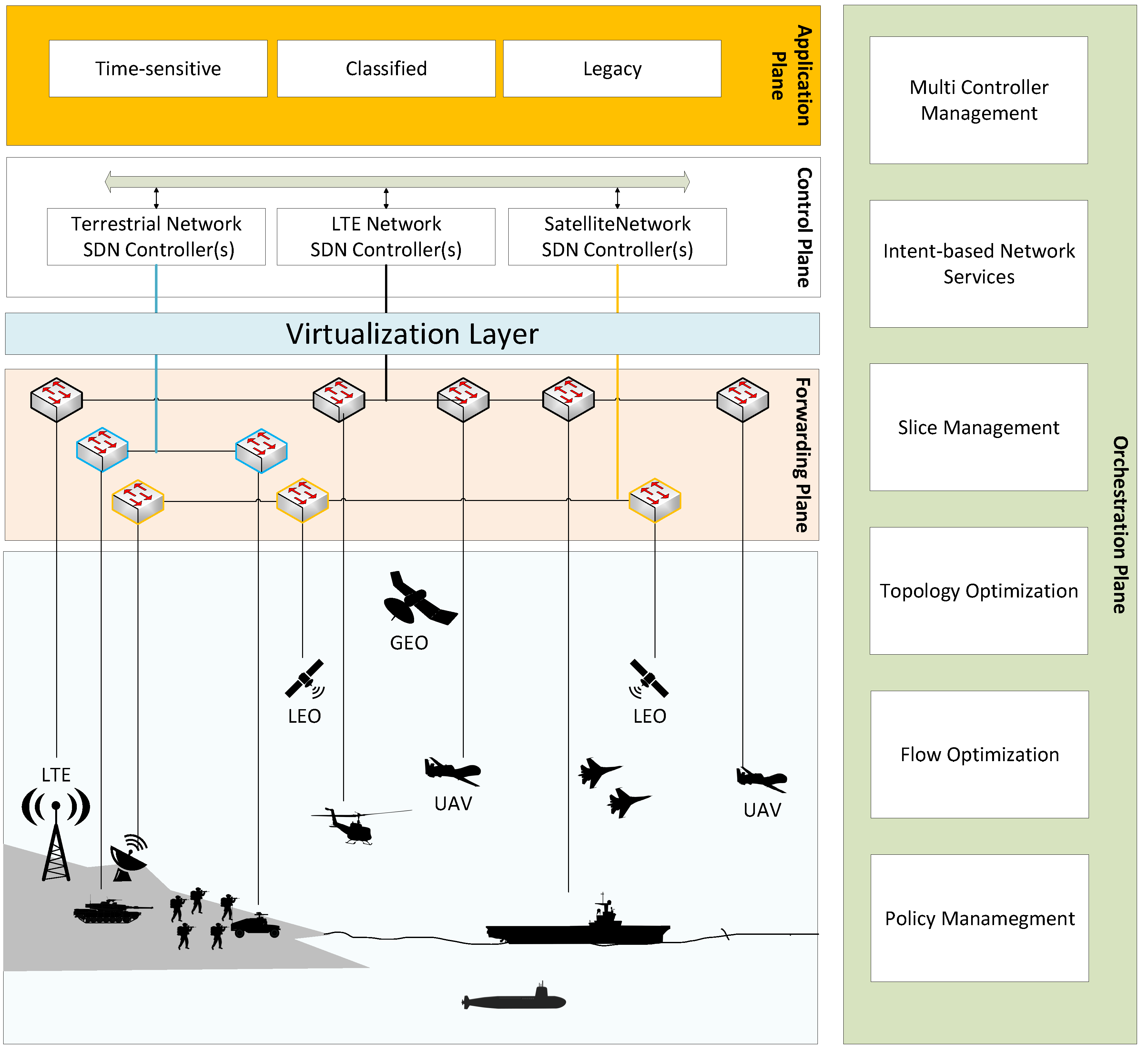}
\caption{SDN layers for MBN-based Military Applications}
\label{Fig:SDN}
\end{figure}
\item \textit{What middleware technologies are suitable for the interoperability of services (distributed application software) in this environment, and why?}
\begin{itemize}
\item[] SDN middleware encapsulates third-party services including databases and application programming interfaces (APIs) that help bridging multiple SDN-enabled systems by going beyond their communication and architectural heterogeneity. Middleware also assists the control plane in interacting with the data plane to perceive the traffic and topology information in a compatible format \cite{Interoperability}. However, in the battlefield context, the attainment of interoperability through middleware is complicated because of the involvement a large number of entities seeking consistent protocol translation and resource discovery support from the middleware. They also increase the management overhead of middleware. Therefore, it is important to select appropriate interoperable technology based on the application requirements and underlying protocols so that the responsiveness and performance of the middleware do not degrade. 
\end{itemize}
\end{enumerate}
\par In literature, there exists a notable number of works that focus on addressing these challenges through efficient SDN orchestration. This paper aims at categorizing and reviewing them in a systematic manner. It also exploits the detailed scope for further research in this direction by exploiting the current research gaps. The major contributions of this paper are listed below. 
\begin{itemize}
\item Proposes a system model and a taxonomy for SDN orchestration, especially from the perspective of tactical networks.
\item Reviews the existing literature on SDN-enabled tactical networks and identifies their pros and cons. 
\item Investigates the current research gaps in augmenting SDN with tactical networks and offers future directions for further improvement in this domain.
\end{itemize}
The rest of the paper is organized as follows: Section \ref{Sec:taxonomy} highlights the proposed taxonomy. The literature review is presented in Section \ref{Sec:LitStrat} to \ref{Sec:LitEnd}. Section \ref{Sec:Future} discusses the research gaps and future directions. Finally, Section \ref{Sec:Done} concludes the paper.
\section{System Model and Taxonomy} \label{Sec:taxonomy}
To simplify the synthesis of different military devices, tactical network and applications, we propose a layered SDN framework as depicted in Fig. \ref{Fig:SDN}. The framework is composed of four planes: application, control, forwarding, and orchestration. Applications with varying QoS and security requirements lie in the application plane. These can be SDN-aware applications communicating directly with an SDN controller, or legacy applications simply sending data through the network. The control plane is composed of multiple, specialized SDN controllers that have the ability to communicate, either in a peer-to-peer fashion or through an orchestrating controller with a global, multi-network view. The forwarding plane consists of networking nodes that have the ability of forwarding packets based on the routing policies implemented by the SDN controllers. Finally, the orchestration plane spans across all layers and is responsible for monitoring and aggregating data to be used in a meaningful way to support efficient network orchestration in terms of controller management, service resiliency, interoperability, and policy enforcement. 
\begin{landscape}
{
\begin{figure}[!t]
\centering 
\captionsetup{justification=centering}
\includegraphics[width=1.35\textwidth]{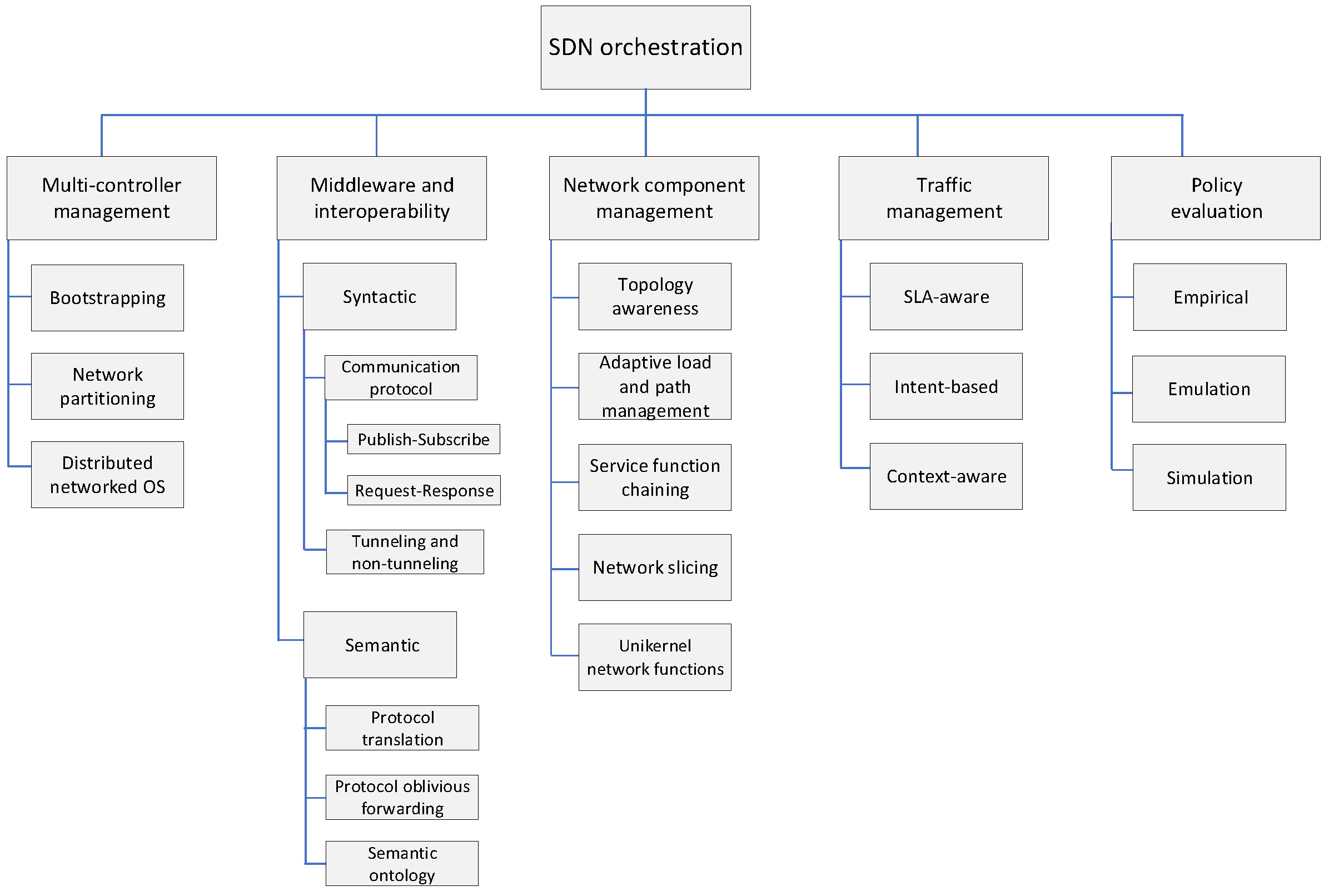}
\caption{A taxonomy on SDN orchestration}
\label{Fig:Taxonomy}
\end{figure}
}
\end{landscape}
\par According to the proposed system model, the policy-driven management of orchestration plane is very essential to enhance the competency of SDN-enabled tactical networks in supporting diverse physical and logical networking components and military applications. In existing literature, accrediting this necessity various SDN orchestration policies has been developed. Fig. \ref{Fig:Taxonomy} depicts a taxonomy on different aspects of SDN orchestration, especially from the perspective of tactical network. In the following Section \ref{Sec:LitStrat}-\ref{Sec:LitEnd}, the detailed description of the taxonomy and its mapping to the existing literature are provided.  
\section{Multi-controller Management}\label{Sec:LitStrat}
\begin{figure}[!t]
\centering 
\captionsetup{justification=centering}
\includegraphics[width=\textwidth]{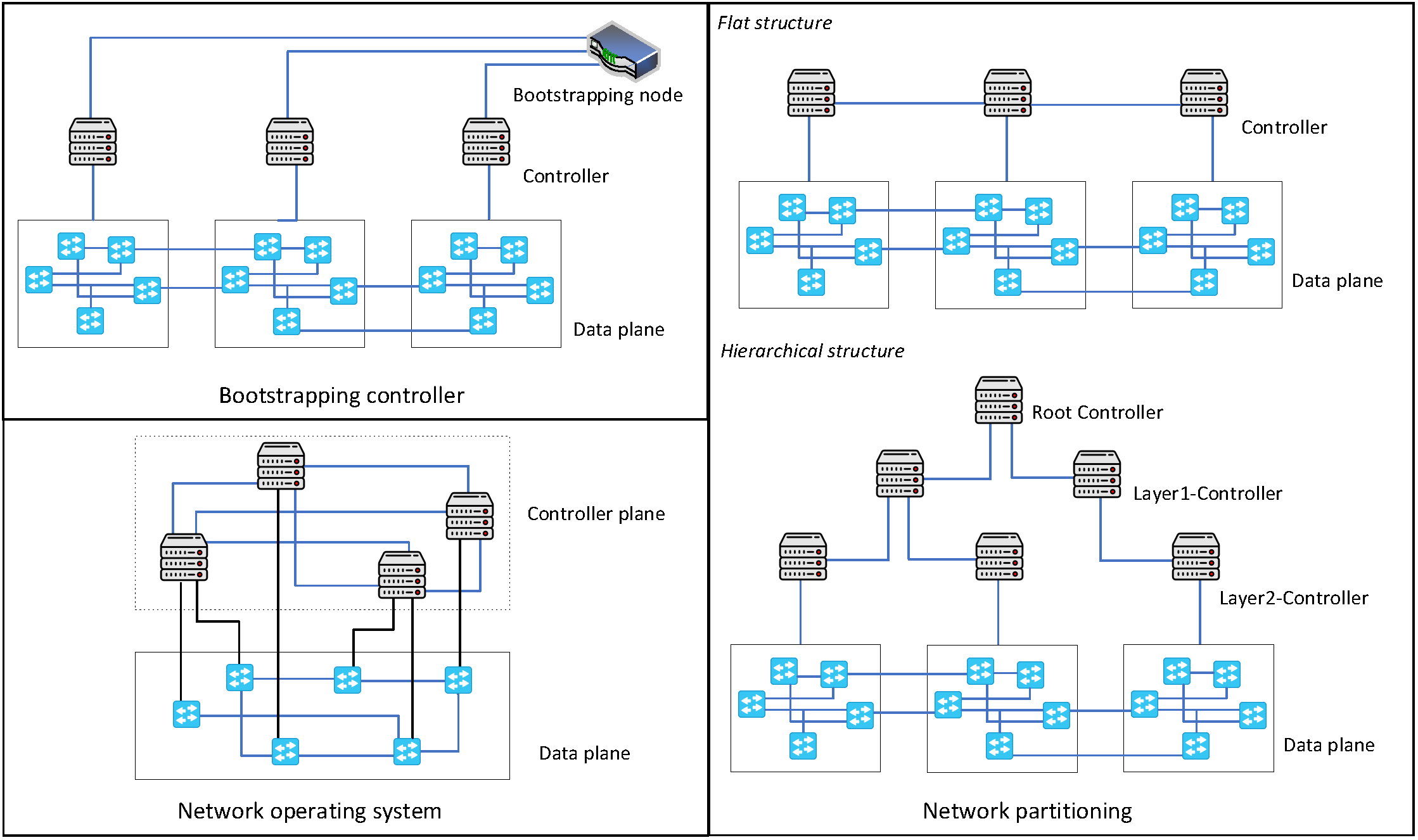}
\caption{Different controller management approaches}
\label{Fig:Controller}
\end{figure}

The implementation of SDN with single controller is unsuitable to deal with the increasing rate of traffic transmission in the battlefields. Moreover, in the tactical context, two military devices such as a submarine and a drone interacting with each other may not be located at the same network domain. In such cases, the implementation of SDN with multiple controllers can play a vital role. The coexistence and collaboration of multi-controllers solve the problems encountered by a single controller and help in cross-domain interactions. However, the operations of multiple SDN controllers in military oriented MBN is subjected to consistency and load balancing-related issues. Three types of controller management approaches (as shown in Fig. \ref{Fig:Controller}) are widely used to deal with these issues in SDN.
\subsection{Bootstrapping}
In bootstrapping, a rendezvous node deploys multiple SDN controllers between the application and the data plane. The bootstrapping node notifies the network configuration information to the controllers, sets their initial topology, and determines the coordination mechanism. To build the topology model for the SDN controllers, the bootstrapping node transmits Link Layer Discovery Protocol (LLDP) packets to various networking nodes including switches and gateways, and substrates the network based on their responses. The bootstrapping node also installs default flow-rules for the data plane so that the network can remain functional even after the failure of the controllers. Moreover, it is capable of increasing or decreasing the number of controllers dynamically according to the requirements of SDN operations.

To simplify the initialization phase of the SDN network, a bootstrapping approach named InitSDN is proposed in \cite{1}. InitSDN helps in modularizing the network applications and facilitates controller migration by only updating their topology. In \cite{2}, another bootstrapping approach is proposed that assists tactical networks to transmit the control commands and the data traffic using the same underlay network. It enables a data plane node to \textit{(i)} identify and register with any of the available SDN controllers, \textit{(ii)} parse the corresponding data flow rules through intermediate switches, \textit{(iii)} initiate a secure control channel with the controller, and \textit{(iv)} interact with the topology database.

Bootstrapping is supportive for dynamic network extension and legacy routing, and can effectively handle uncertain failures within the control and data plane \cite{3}. A bootstrapping networking device can also serve the purpose of a edge computing node. However, for bootstrapping, the controllers and data plane nodes are required to be explicitly accessible, which is not recommended for military use cases. Moreover, bootstrapping a wireless SDN is a challenging task as the controllers and data plane nodes only share local connectivity information and resist the attainment of global bootstrapping convergence instantly.

\subsection{Network Partitioning}
In network partitioning, the data plane is divided into multiple domains, and for each domain a local SDN controller is assigned. The interactions between the controllers are made through either a hierarchical or a flat structure. In a hierarchical structure, a group of controllers residing at the upper layer explicitly manage the controllers in the immediate underneath layer. The number of these logical layers is set by the network operator based on the network topology size, the traffic load, and the network resource availability. Moreover, in this setup, the controllers in the same layer do not communicate with each other directly. Their internal communication happens via the upper layer controllers. Conversely, in the flat structure, the controllers of various data plane domains spontaneously interact with each other using east and west bound APIs to maintain a global view of the underlying network. Among the celebrated SDN controllers, ONIX, HyperFlow and OpenDayLight use the flat structure whereas, Kandoo, Orion and D-SDN follow the hierarchical structure \cite{4}. 

Nevertheless, network partitioning becomes vigorous when the traffic load is evenly distributed among the controller. By exploiting the k-means clustering algorithm and the cooperative game theory, a load management policy for multi-controllers is proposed in \cite{5}. The policy enables a data plane node to form coalitions with other nodes and balance the topology size for each controller in partitioned SDN. Internet 2 OS3E and Internet Topology Zoo is used to evaluate the performance of the policy. On the other hand, in \cite{6}, a Louvain heuristic algorithm is developed to limit the number of data plane nodes managed by a controller so that the controllers do not get overloaded.

Network partitioning is supportive to wireless networking because of its localized characteristics and inherently complements the realization of edge computing. However, the interaction of two controllers in partitioned networks is time consuming as it requires the assistance of multiple intermediate controllers. The impact of such delays in tactical scenarios is evaluated in \cite{7}. Moreover, in partitioned networks, a significant amount of resources is consumed only to synchronize controllers, which is not suitable for resource constrained environments like battlefields.
\subsection{Networked Operating System (NOS)}
In this approach, a physically distributed but logically centralized network operating system runs across multiple controllers. The network applications within the operating system support the controllers to handle the traffic flow and maintain a global view of the network \cite{8}. Additionally, these applications can enable any data plane node to connect with different controllers but allow only one controller to manage that node at a time. If the controller fails, another controller is set as the node manager based on a consensus-based leader selection algorithm. Moreover, the operating system supports the dynamic updates of the applications without interrupting the traffic flow. SDN frameworks including Open Network Operating System (ONOS), Switch Light, Open Network Linux (ONL), DENT and Coriant predominantly follow the concepts of a networked operating system in their control plane implementation \cite{9}.

Apart from the benchmarks, there exist several customized implementation of network operating system for SDN controllers. For example, in \cite{10}, a network operating system named MNOS is developed that augments the cyberspace to mimic defence technique and protects the controllers from data alteration. It also creates the functional equivalent variants of the controllers using dissimilar redundancy design principles to overcome their device-level heterogeneity. In \cite{11}, another network operating system named NOSArmor is proposed that augments security blocks to the controllers. The blocks are responsible for role-based authorization, location tracking, link verification, rule-based negotiation, protocol verification, system call checking and resource management. Moreover, there are some extensions of network operating system that either protect the control plane from the compromised controllers by exploiting the packet trajectory information \cite{12} or apply lightweight virtualization techniques such as containers for resource constrained controllers \cite{13}. 
  
Network operating systems are modular and fault tolerant. Additionally, the expansion and consolidation of network operating system-based control planes are comparatively easier and less time consuming. However, such control planes are required to be deployed locally for synchronization, which may not be feasible for military use cases requiring cross-network domain communications. They also lack support for channel-level management of MBN \cite{14}.
\section{Middleware and Interoperability}
To ensure efficient tactical interactions, SDN middleware requires to support interoperability between the control and the data plane nodes. The overall interoperability of any system can be discussed from two perspectives, syntactic and semantic. Table \ref{Tab:SynSem} illustrates the differences between syntactic and semantic interoperability. In the literature, there are different techniques that help in enabling syntactic and semantic interoperability in SDN. However, these interoperability techniques have their own pros and cons in dealing with the dynamics of battlefield communications and diverse traffic priorities.
\begin{table}[!t]
\centering 
\small
\caption{Differences between syntactic and semantic interoperability}\label{Tab:SynSem} 
\begin{tabular}{|p{1.5 cm}|p{4 cm}|p{4 cm}|}
\hline
Facts & Syntactic interoperability & Semantic interoperability \\\hline
Targets	& Data exchange	& Data interpretation \\\hline
Deals with & Format of data	 & Contents and attributes of data \\\hline
Enablers & Communication protocol & Information model \\\hline
\end{tabular}  
\end{table}   
\subsection{Syntactic}
Syntactic interoperability is responsible for the synergies of the data packets and their formats transmitted and packaged by the heterogeneous control and data plane nodes. It is also regarded as the prerequisite for attaining semantic interoperability in SDN. Syntactic interoperability explicitly depends on the communication protocols offered by the middleware and the characteristics of the overlays that logically connects the nodes with the middleware. Different communication protocols and overlay mechanisms associated with the syntactic interoperability are discussed below.
\subsubsection{Communication Protocols}
Most of the existing SDN middleware systems have a message-oriented architecture that allows them to handle uncertain communication delays during interactions with different control and data plane nodes. Additionally, the functionalities of a message-oriented middleware are highly scalable compared to that of a remote procedure call-based middleware \cite{15}. Two types of communication protocols such as Publish-Subscribe (PubSub) and Request-Response (RR) are widely used in message-oriented systems. 
\begin{itemize}
\item[i.] \textbf{Publish-Subscribe}: PubSub communication protocols assist the control plane node in publishing the commands to the middleware and enable data plane nodes to get the respective commands from the middleware. The opposite happens when data is transferred from the data plane to the control plane. PubSub protocols support event-driven interactions between the communicating entities. Message Queuing Telemetry Transport (MQTT), Data Distribution Service (DDS) and Advanced Message Queuing Protocol (AMPQ) are among the most used PubSub communication protocols.
\begin{itemize}
\item[$\bullet$] \textbf{\textit{Message Queuing Telemetry Transport (MQTT)}}: MQTT protocol defines a MQTT broker at the middleware and a set of logical clients over the control and data plane to publish and subscribe information. MQTT sorts information in topics and allow nodes to subscribe multiple topics and receive all information published under each topic. For example, in \cite{16}, an MQTT enabled SDN framework for UAV swarms is proposed that creates different MQTT information topics for exchanging network conditions, security policies, QoS requirements, electronic state and controller commands. 

Usually MQTT depends on TCP for data transmission. There is a variant of MQTT for sensor networks, named MQTT-SN that uses either UDP or Bluetooth for transmitting data. MQTT is also used to create multicast trees between the publishers and the subscribers for minimizing data transfer delay \cite{17}. In another work, MQTT has been exploited in multi layers to offer network interoperability for the controllers deployed in hierarchical structure \cite{18}. 

MQTT is considered highly feasible for Internet of Things-driven interactions because of its lightweight structure and minimized data packets \cite{19}. Nevertheless, MQTT often experiences serious traffic congestion problem at the broker side and requires Transport Layer Security (TLS) support. Moreover, MQTT is less resilient to the mobility of subscribing and publishing nodes, and prone to single point failure. These limitations can resist the real timeliness of the system and increase overhead of the middleware \cite{20}.  
 
\item[$\bullet$] \textbf{\textit{Data Distribution Service (DDS)}}: DDS allows asynchronous data exchange among communicating entities without implementing any logical broker. Unlike MQTT, DDS incorporates a built-in discovery mechanism that assists subscribers in finding the available publishers for interactions. The default transport layer protocol for DDS is UDP, although it can be easily integrated with TCP. The header length of DDS is 16 bytes which is 8 times higher than that of MQTT and possesses 20 more QoS levels for controlling volatility, resource utilization, availability, delivery, reliability, ownership, duplication, and latency tolerance of the data. Therefore, a DDS middleware requires to extract the data-centric information of the packets for their QoS-satisfied distribution to the subscribers \cite{21}.

In SDN, the concept of DDS middleware has been widely used to manage the distributed control plane. For example, in \cite{22}, a DDS-based hierarchical controller plane structure is modelled that distributes time-critical synchronization and system breakdown information among the controllers by publishing their type in proactive manner to achieve better performance. Another SDN control mechanism is developed in \cite{23} for dynamically configuring network based on the importance of shared data among the digital twins. The mechanism set this data importance in terms of the latency sensitivity attribute of the packets defined by the DDS QoS level. Moreover, in \cite{24}, a DDS-based SDN middleware is considered that supports on-demand access to UAV-aided services from authorized entities at the ground. It also facilitates distributed DDS orchestration to enhance interoperation and meet mobility constraints of UAVs.

DDS supports security plugin models and offers vendor level interoperability using RTPS (Real Time Publish Subscribe) protocol. Due to built-in QoS maintenance mechanism, DDS also performs better in low latency communication. However, DSS is heavyweight for resource constrained battlefield networking nodes and consumes more bandwidth than MQTT. 

\begin{figure*}[!htbp]
  \centering
  \captionsetup{justification=centering}
  \begin{minipage}[t]{\textwidth}
    \centering
    \captionsetup{justification=centering}
    \vspace{0pt}
    \includegraphics[width=0.6\textwidth]{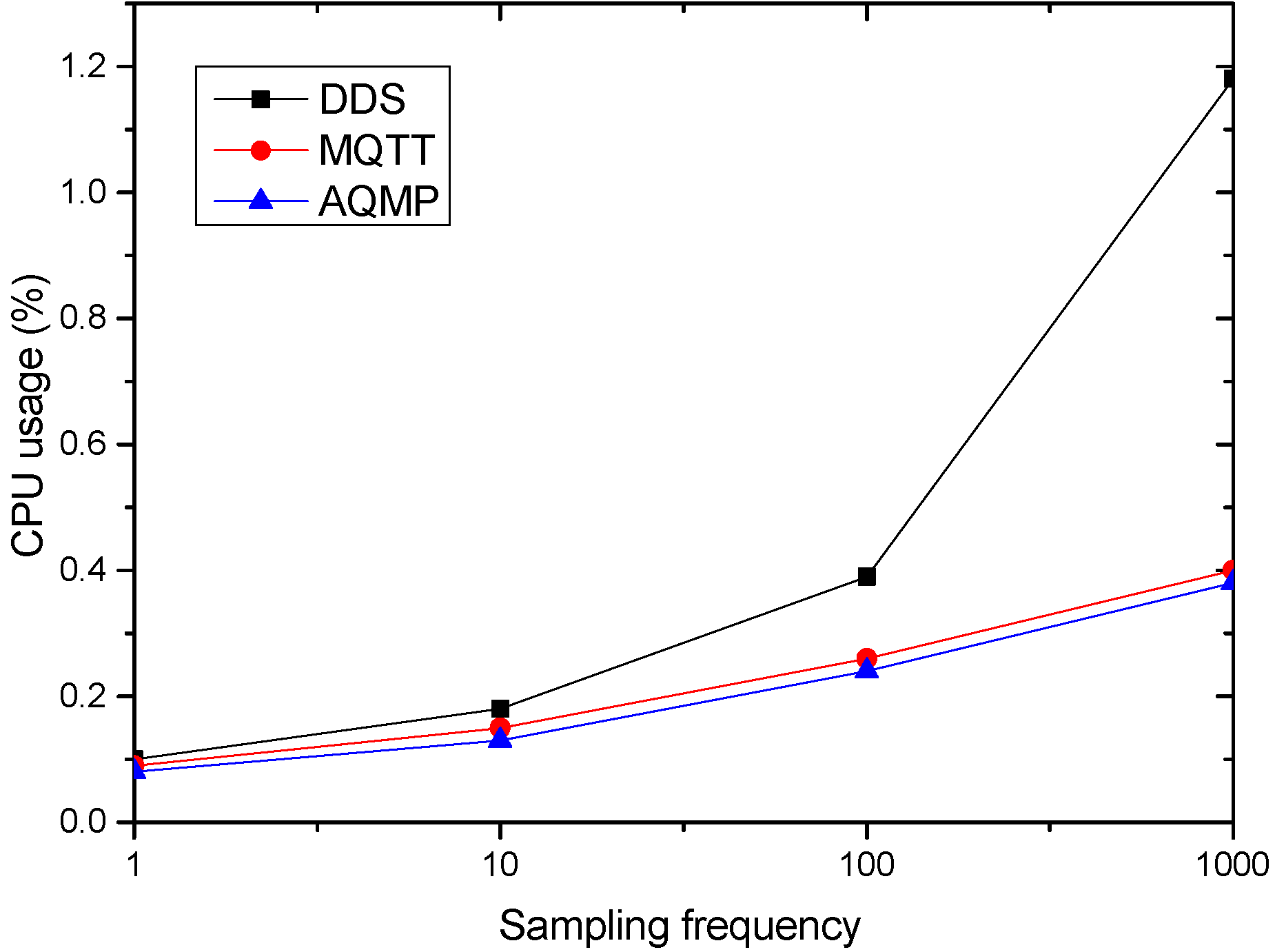}
    \caption*{a. CPU usage}
  \end{minipage}
  \qquad
  \hfill
  \begin{minipage}[t]{\textwidth}
  \centering
  \captionsetup{justification=centering}
    \vspace{0pt}
    \includegraphics[width=0.58\textwidth]{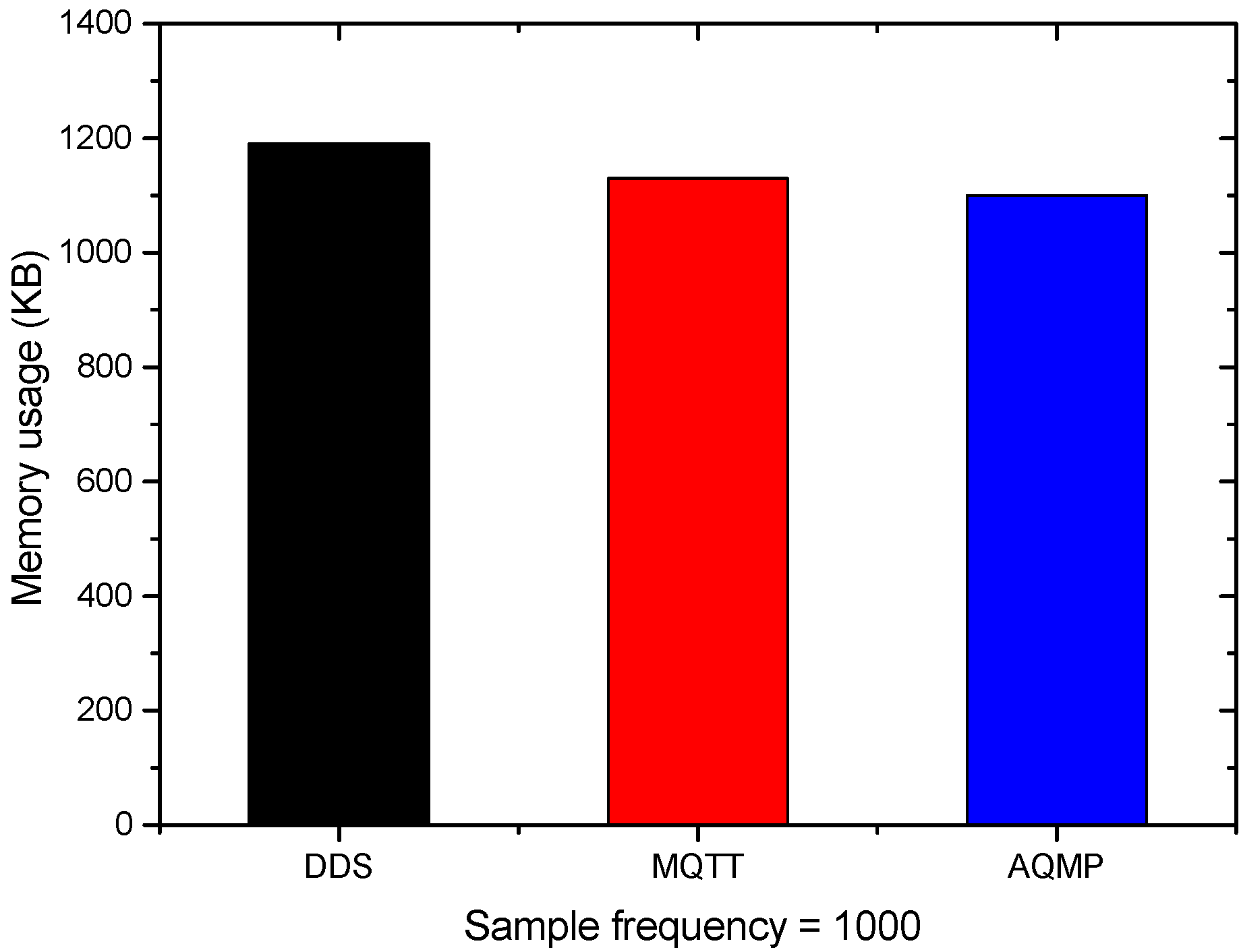}
    \caption*{b. Memory usage}
  \end{minipage}
  \qquad
  \hfill
  \begin{minipage}[t]{\textwidth}
  \centering
  \captionsetup{justification=centering}
  \vspace{0pt}
    \includegraphics[width=0.6\textwidth]{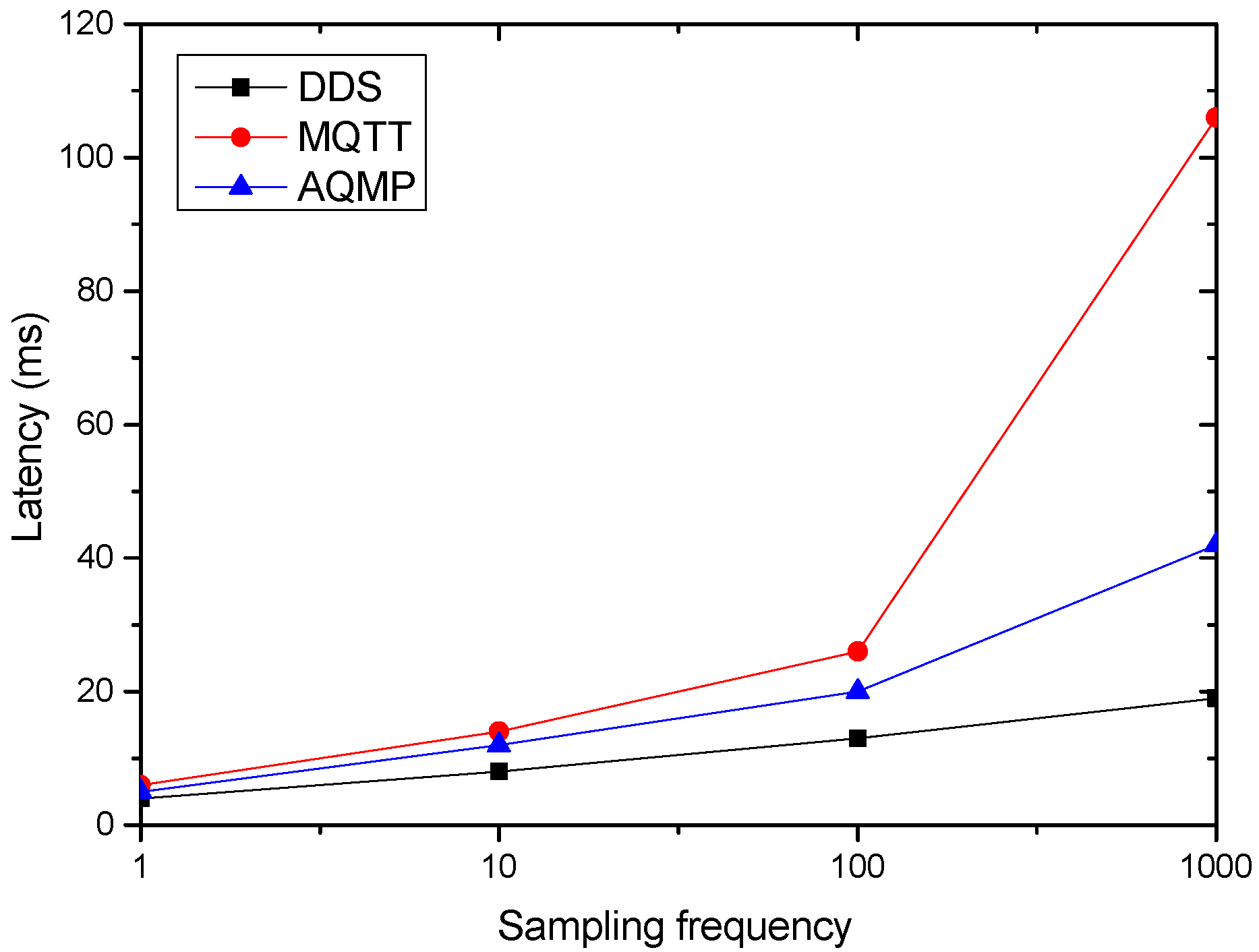}
    \caption*{c. Latency}
  \end{minipage}
%  \vspace{-0.25cm}
\caption{Comparison between MQTT, DSS and AMQP \cite{28}}
\label{Fig:PuBSub}
\end{figure*}

\item[$\bullet$] \textbf{\textit{Advanced Message Queuing Protocol (AMQP)}}: In AMQP broker, the published messages received by the exchange component are organized in multiple queues based on a set of certain rules called bindings. The published messages contain various meta-data that help the broker to retrieve context and priority of the packets without exploiting the payload directly. Similar to MQTT, AMQP exploits TCP for data transmission and provides three QoS levels namely, i. at most once, ii. exactly once and iii. at least once. However, the header length of AMQP is 8 bytes higher than that of MQTT.

AMQP-based SDN middleware systems are often used to build distributed control plane. In \cite{25}, such a middleware has been considered that augments RabbitMQ and ActiveMQ with AMQP for supporting reliable message communication among the controllers. Similarly, in \cite{26}, another AMQP middleware is modelled to exchange information regarding network bandwidth, network topologies and inter-connected nodes among the distributed controllers.

Nevertheless, AMQP helps in enhancing communication flexibility by providing a scope to dynamically integrate different network standards and protocols. Additionally, the AMQP packet size is negotiable that makes it suitable for transferring large number of payloads. On the contrary, AMQP does not facilitate automatic resource discovery like DDS and lacks explicit support to enable Last-Value-Queues update. AMQP can also create a large backlog of messages when there is a poor availability of network resources and resists real-time battlefield communications by increasing the network delay \cite{27}. Additionally, Fig. \ref{Fig:PuBSub} illustrates the differences of MQTT, DDS and AMQP from the perspective of CPU, memory and latency-driven performances. 
\end{itemize} 

\item[ii.] \textbf{Request-Response}: In RR communication protocols, when a data plane node needs any command from the control plane, it sends a request to the corresponding controller through middleware. In response, the controllers transfer necessary instructions to the data plane node. The opposite happens when the control plane seeks state information from the data plane. RR issues both request and response packets in a synchronous manner. In Table \ref{Tab:PubSub}, a summary comparison between PubSub and RR has been illustrated. Constrained Application Protocol (CoAP) is one of the most celebrated RR protocols that deals with IoT communications in resource constrained networking environments \cite{Paola}. 
\begin{table}[!t]
\centering 
\small
\caption{Comparison between PubSub and RR}\label{Tab:PubSub} 
\begin{tabular}{|p{3 cm}|p{4 cm}|p{4 cm}|}
\hline
Facts & PubSub & RR \\\hline
Suitable for	& Competitive, unreliable network	& Robust, reliable network \\\hline
Traffic load	& High	& Low \\\hline
Interaction driver	& Report-by-exception (RBE) & Polling at regular interval \\\hline
Dynamic scaling	& Adaptive	& Inflexible \\\hline
Security augmentation &	Complicated	& Easy \\\hline
\end{tabular}  
\end{table}   
\begin{itemize}
\item[$\bullet$] \textbf{\textit{Constrained Application Protocol (CoAP)}}: CoAP relies on both UDP and RESTful protocol that makes it more compatible for resource constrained IoT devices. Moreover, CoAP offers reduced implementation and communication complexities compared to other RR protocols like HTTP.  As a means of reliability. CoAP also incorporates an exponential back-off feature-based retransmission mechanism. CoAP supports two different levels of QoS functionalities, namely \textit(1) Confirmable, \textit(2) Non-Confirmable. Its header length is 4 bytes and can be easily augmented with cellular networks.

In the literature, there exist several researches studies where CoAP has been used to model communications among distributed control plane entities. For example, in \cite{29}, a control plane structure for software defined wireless network is developed that exploits CoAP for exchanging topology discovery and flow control information among the controllers. In another work \cite{30}, CoAP has been used to allow controllers for managing flow tables, modifying node routing characteristics, and obtaining data plane information with respect to link quality, geographical location and energy level. Moreover, in \cite{31}, a real-world SDN middleware named Ride has been developed that exchanges CoAP packets for managing a workflow consisting various tasks including host registration, network configuration, on-demand network state analysis, fault detection and recovery. CoAP offers faster wake up times and extended sleepy states that consequently improves energy consumptions of control and data plane nodes. However, CoAP has limitations in communicating devices using Network Address Translation (NAT) technique.
\end{itemize}
%
%main item 
\end{itemize}
\subsubsection{Tunneling and non-tunneling}
Tunneling allows private communications to exchange data packets across a public network using encapsulation. By default, it supports encryption and helps in establishing secure and remote connections among the networks. These features make tunneling highly feasible to use in virtual networks. There exist different tunneling protocols such as Virtual Extensible LAN (VXLAN), GPRS Tunneling Protocol (GTP), Network Virtualization using Generic Encapsulation (NVGRE), stateless transport tunneling (STT) and Network Virtualization Overlays 3 (NVO3) that simplifies the realization of virtual networks \cite{32}. Moreover, in SDN, tunneling is often used to manage connection among the data plane nodes, especially during the uncertain mobility of packet destinations \cite{33}. In such cases, tunnels are created dynamically to handover data packets from the previous serving switch to the current serving switch of the destination node. On the other hand, in \cite{34}, an SDN-enabled dynamic multipath forwarding technique has been developed that can merge traffics of multiple tunnels at any data plane node based on source-destination address with a view to minimizing the number of flow entries within the system. 

Moreover, there exist other initiatives that focus on improving tunneling mechanisms in SDN. For example, in \cite{35} a Match-Action Table (MAT) programming model-based IP tunnel mechanism, named MAT tunnel is developed that allows controllers to set flow table entries with both encapsulation and decapsulation specifications of the corresponding tunnel. It consequently reduces the overhead of manually configuring the tunnel interface at the data plane. Similarly, in \cite{36}, another tunneling mechanism is developed that detects multiple shorter repair paths when a single link failure happens in SDN. This feature helps in faster fault recovery.

However, the packet drop rate in tunneling increases unevenly when mixed traffic (voice and video) are transferred. Forward error correction in this case incurs additional bandwidth overhead and wastes network capacity. The repackaging feature of tunneling reduces the effective size of data packets and affects the transfer delay. It consequently increases packet fragmentation that consumes additional memory and processing power at the destination node for merging. Because of these limitations, tunneling is often discouraged while transferring large amounts of data to resource constrained destinations. Therefore, non-tunneling communications for virtual networking is gradually getting attention in both research and industry. In \cite{37}, a non-tunneling protocol named FlowLAN is developed that adopts Network Prefix Translation technique to augment both the physical and logical addresses of packet destination nodes and tags them in the flow field of the packet header with respect to the corresponding network identifier. It helps realizing the virtual networks as a distributed system that can communicate without encapsulation or decapsulation. To support the movement of cells in LTE network, another non-tunneling approach named MocLis is developed in \cite{38}. MocLis adopts Locator/ID split approach while dealing with the mobility of cells and their nested user equipment. Nevertheless, non-tunneling approaches lack standardization that makes them less compatible to apply in highly heterogeneous communication environments like battlefield. 
\subsection{Semantic}
In SDN, a middleware needs to support semantic interoperability to ensure the unambiguous interpretation of command and status information that is exchanged between the controllers and data plane nodes. It simplifies the knowledge discovery between these two planes. Semantic interoperability acts as a function of semantic interoperability and fails drastically if the data packets are distorted during transmission from source to destination. There exist different techniques including protocol translation, protocol oblivious forwarding and semantic ontology that enable semantic interoperability in SDN. 
\subsubsection{Protocol Translation}
\begin{figure}[!t]
\centering 
\captionsetup{justification=centering}
\includegraphics[width=\textwidth]{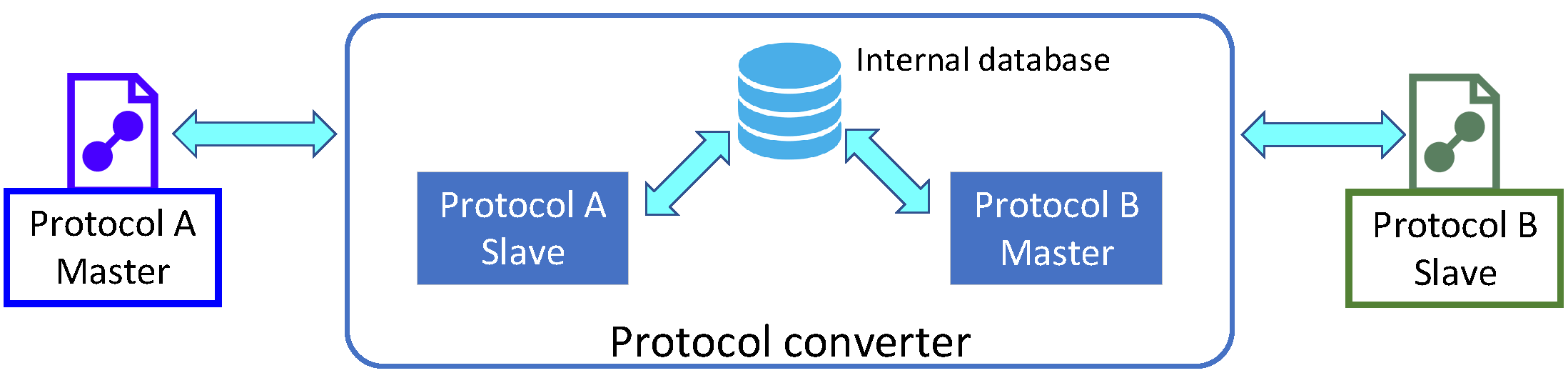}
\caption{Architecture of a protocol converter}
\label{Fig:PT}
\end{figure}

Protocol translation converts the data, commands and time synchronization information issued by the control plane into the compatible format of the data plane nodes in which they are navigating. It also enables the data plane nodes to interact with controllers despite of the differences in their native protocol stacks. To perform this operation, a Protocol Converter software installed on the middleware removes the protocol headers of the sender completely and wrap the payload with the target protocol header \cite{39}. There are different technical companies like Cisco and Valin corporation that develop software solutions for protocol translation. Fig. \ref{Fig:PT} depicts the internal architecture of a conceptual protocol converter software.  

In \cite{40}, the operations of a protocol translating middleware named TableVisor is discussed.  TableVisor uses the match-action architecture to match the intents of the exchanged data packets to the existing flow table entries, action space and target header fields. The expressiveness of TableVisor is translating protocols is defined by the intersections of possible command attributes from both source and target protocol. The protocol translation mechanism discussed in \cite{41} shows almost the similar functionalities like TableVisor. However, for \cite{41}, the translation rules are defined by the controllers, not by the middleware. Conversely, in \cite{42}, the middleware translates the source data and protocol commands into multiple segments as per the primitive network requirements with respect to latency, packet collision and packet delivery rate so that the destination nodes can easily parse the segments with their default protocol stack and set the rank for each requirement. 

Although protocol translation helps in alleviating protocol and data format-wise heterogeneity of control and data plane nodes, it limits the scope of simultaneous interactions. It requires an in-depth understanding of the packets that urge to deploy trusted middleware systems across the network. However, such facilities are not often possible to ensure in constrained communication environments like battlefield. 
\subsubsection{Protocol Oblivious Forwarding}
Protocol oblivious forwarding makes the format of a packet transparent to the data plane nodes. In this case, the data plane nodes extract and assemble key features from the packet header to conduct flow table lookups based on the controller instructions. It enables data plane to support any new protocols and forwarding requirements in a flexible manner. To perform this operation, packet meta-data are augmented with generic information including flow logic and life span. The difference between protocol translation and oblivious forwarding is illustrated in Table \ref{Tab:TransFor}. 

A protocol-oblivious forwarding-based routing mechanism is proposed in \cite{43} that can redirect a packet to multiple destination addresses in a multi-homing scenario. It completements the SDN ability of switching transmission path dynamically and enables the destinations to adjust packet receiving rate as per the status of network resources. Moreover, to assist protocol oblivious forwarding in perceiving device-level context, a State Parameter Field is augmented to its generic structure in \cite{44}. It also incorporates a direct entry matching policy for flow table lookup that enables protocol oblivious forwarding to check device status in time optimized manner. Moreover, in \cite{45}, the concept of protocol oblivious forwarding has been extended to offer protocol independent interactions among the controllers arranged in a hierarchical structure. It enhances the flexibility in distributed controller operations. 

Despite having certain advantages over protocol translation, protocol oblivious forwarding is considered infeasible to sensitive communications as it lacks explicit security measures. Therefore, to protect the protocol oblivious forwarding operations from diverse attacks, a proactive security framework for SDN is proposed in \cite{46}. Moreover, protocol oblivious forwarding depends on a set of stateful information which makes it less resilient to failure or alteration of the networking system. 

\begin{table}[!t]
\centering 
\small
\caption{Differences between Protocol translation and oblivious forwarding}\label{Tab:TransFor} 
\begin{tabular}{|p{5.5 cm}|p{5.5 cm}|}
\hline 
Protocol translation & Protocol oblivious forwarding \\\hline
Requires protocol specific knowledge & Protocol specific knowledge is oblivious \\\hline
Parsing packet data for target protocol is difficult in real-time & Extraction of meta data from packet is easier \\\hline
Conversion or translation support for user-defined or newly introduced protocols are not always available & Data plane can adopt any protocols \\\hline 
\end{tabular}  
\end{table}   

\subsubsection{Semantic Ontology}

\begin{figure}[!t]
\centering 
\captionsetup{justification=centering}
\includegraphics[width=\textwidth]{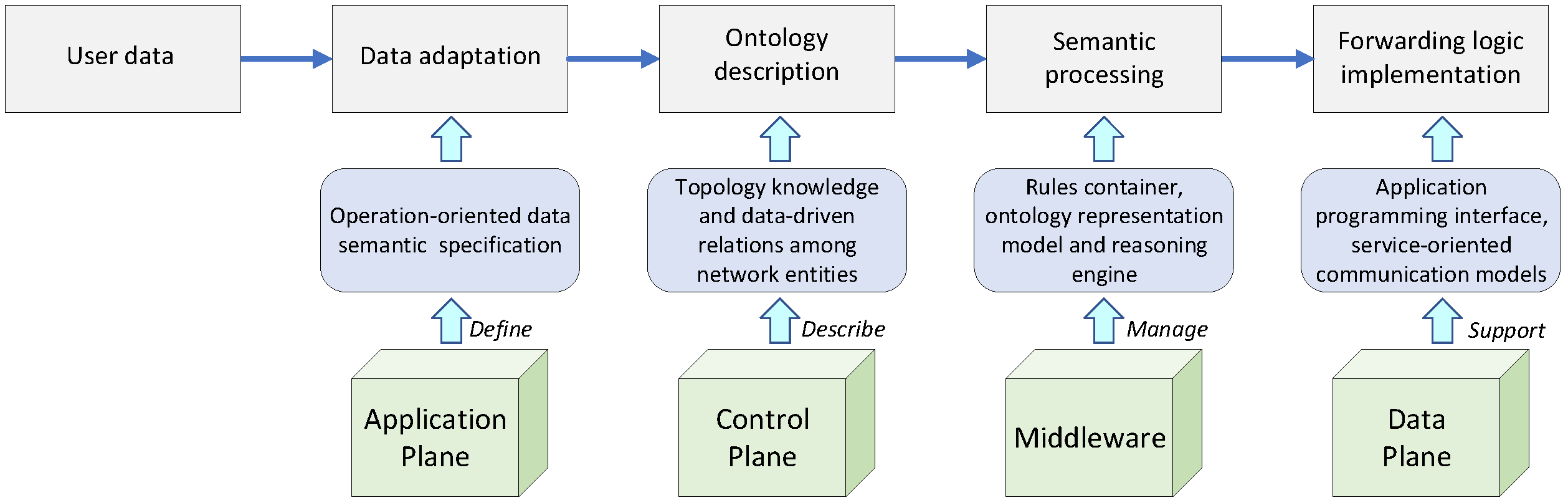}
\caption{Semantic ontology-based operations in SDN layers \cite{48}}
\label{Fig:Ontology}
\end{figure}

A significant amount of control data is exchanged between control and data plane nodes while transferring network packets from a place to another. The existing Network Operating System (NOS)-based control data modelling techniques such as type checking and code templating perform well when the flow rules are static. To parse the non-deterministic behaviors of applications and networks in the flow rules and modelling the control data accordingly, semantic ontology is often used. Semantic ontology incorporates various reasoning rules and integrity constraints that helps in automating state inference across the SDN layers. Additionally, it simplifies the remote configurations of data plane nodes and allow controllers to define complex data relationships \cite{47}. An illustration of semantic ontology-based operations in SDN domain is depicted in Fig. \ref{Fig:Ontology}. 

Based on the concept of semantic ontology, an autonomous fault management agent for SDN is developed in \cite{49}. It compares network status with semantic models using Bayesian reasoning as inference method for determining the category of a fault. In another work \cite{50}, sematic ontology has been applied to automate the creation of virtual network functions (VNFs). It also fosters the synthesis of VNFs with user requirements and enabled controllers to recommend similar services based on network service description (NSD). Moreover, in \cite{51}, another semantic-based framework for distributed control plane is proposed that incorporates local ontology from each controller and forwards them to the master controller for ensuring overall semantic interoperability within the network.  

However, the scope of applying semantic ontology is constrained as it depends on specific format of data and all entities within the network should have in-depth understanding of that format. Moreover, semantic ontology can expose data to security threats for the sake of reasoning which is not acceptable during battlefield communications. 

\section{Network Component Management}
Conceptually, network components are classified into two categories, network infrastructure and network services. Network infrastructure incorporates the topology and the data forwarding paths. From the perspective of SDN, network slices can also be considered as a virtualized infrastructure for the network. Conversely, network services provide support for caching, network address translation, encryption, and intrusion detection. Recently network services are set to be decupled from proprietary hardware to virtualized software platforms using Network Function Virtualization (NFV) techniques. Although it is not a must to implement SDN and NFV together, both technologies can complement each other in enhancing network automation. For example, the implementation of SDN without virtualizing network functions results in hardware dependency which is conflicting with the instinct of SDN that focuses on performing network control through software. In this part of the report, existing approaches to manage network components are discussed in an integrated manner. Section \ref{sec:sub1start}-\ref{sec:sub1end} discuss the approaches from the perspective of network infrastructure whereas section \ref{sec:sub2start}-\ref{sec:sub2end} focus on the approaches based on network service. 
\subsection{Topology Awareness} \label{sec:sub1start}
\begin{figure}[!t]
\centering 
\captionsetup{justification=centering}
\includegraphics[width=\textwidth]{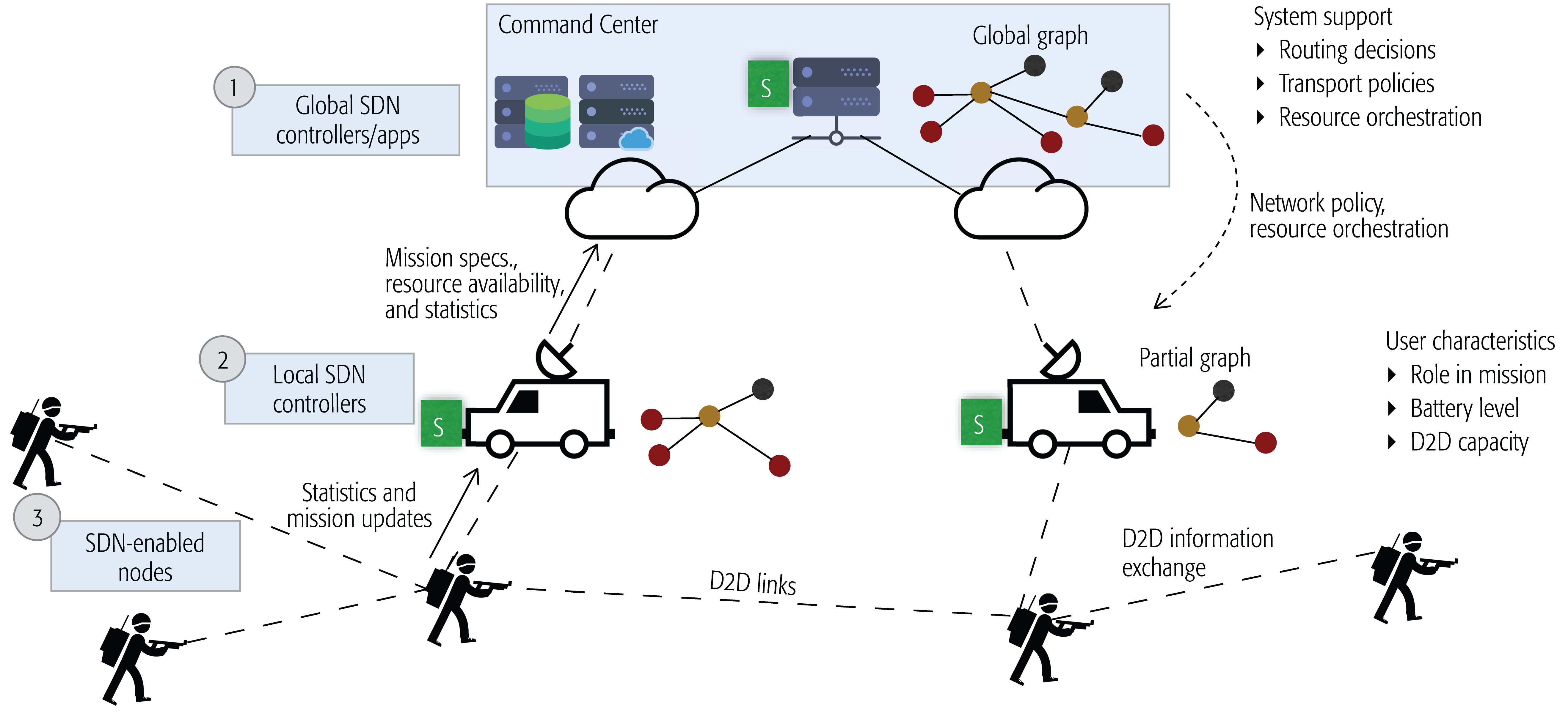}
\caption{SDN-enabled MANET for battlefield communication \cite{52}}
\label{Fig:Manet}
\end{figure}

As noted, tactical operations often take place in inaccessible locations where the arrangement of infrastructure network is difficult. In such cases, on-demand network services can be offered by creating MANET. MANET enables the participating nodes to interact with each other with the goal of completing their assigned tasks. Moreover, MANET provides a scope to integrate the concept of SDN for efficiently coordinating the communicating nodes in pursuing their collective goal. An SDN-enabled MANET structure for battlefield communication is depicted in Fig. \ref{Fig:Manet}. However, the network topology in MANET embraces complex configurations and can change very frequently. Therefore, from the perspective of tactical operations relying on Mobile Ad-hoc Network (MANET), topology awareness is very important. Topology awareness refers to the complete understanding of various dynamics related to the communicating entities and their underlying network while making any network management decision. It consequently helps in optimizing the packet routing path, consolidating the number of redundant networking nodes, scaling-up the network, and deploying edge computing nodes.

In literature, there exists a notable number of works that address the topology awareness in SDN-enabled MANET. For example, a distributed SDN controller placement problem for MANET is formulated in \cite{53}. This work explicitly considers the topology of the network in terms of controller’s accessibility from the data plane nodes and minimizes the cost of circulating synchronization messages among the controllers within the topology. In another work \cite{54}, the communication and topology-driven incompatibility between SDN (inherently centralized and structured) and MANET (inherently distributed and dynamic) is discussed. It also develops a protocol for localized data plane nodes that dynamically adapts the packet routing path according to the changes in network topology without solely relying on the centralized SDN controllers. The performance of the developed protocol is validated using a real-world dataset mentioned in \cite{55}. Furthermore, a multi-path transmission control protocol for decreasing network handover delay and improving transmission throughput in SDN-enabled naval battlefield network is proposed in \cite{56}. The ad-hoc network model also incorporates a connectable relay point to maintain the communications during uncertain topology changes. On the other hand, to ensure security in SDN-enabled MANET during topology alteration, a distributed firewall system is developed in \cite{57}. It relies on ONOS control platform and control the access of unreliable ad-hoc nodes by distributing filter rules across the network. Similarly, in \cite{58}, a flow-based framework for tactical mobile ad-hoc network is proposed that exploits both machine learning-based classification and SDN concepts for anomaly detection within the network topology. However, these topology-aware solutions are very less-adaptive and scalable to deal unpredictable growth of packets in different bearer channels of tactical ad-hoc network. 

\subsection{Adaptive Load and Path Management}

\begin{figure}[!t]
\centering 
\captionsetup{justification=centering}
\includegraphics[width=.50\textwidth]{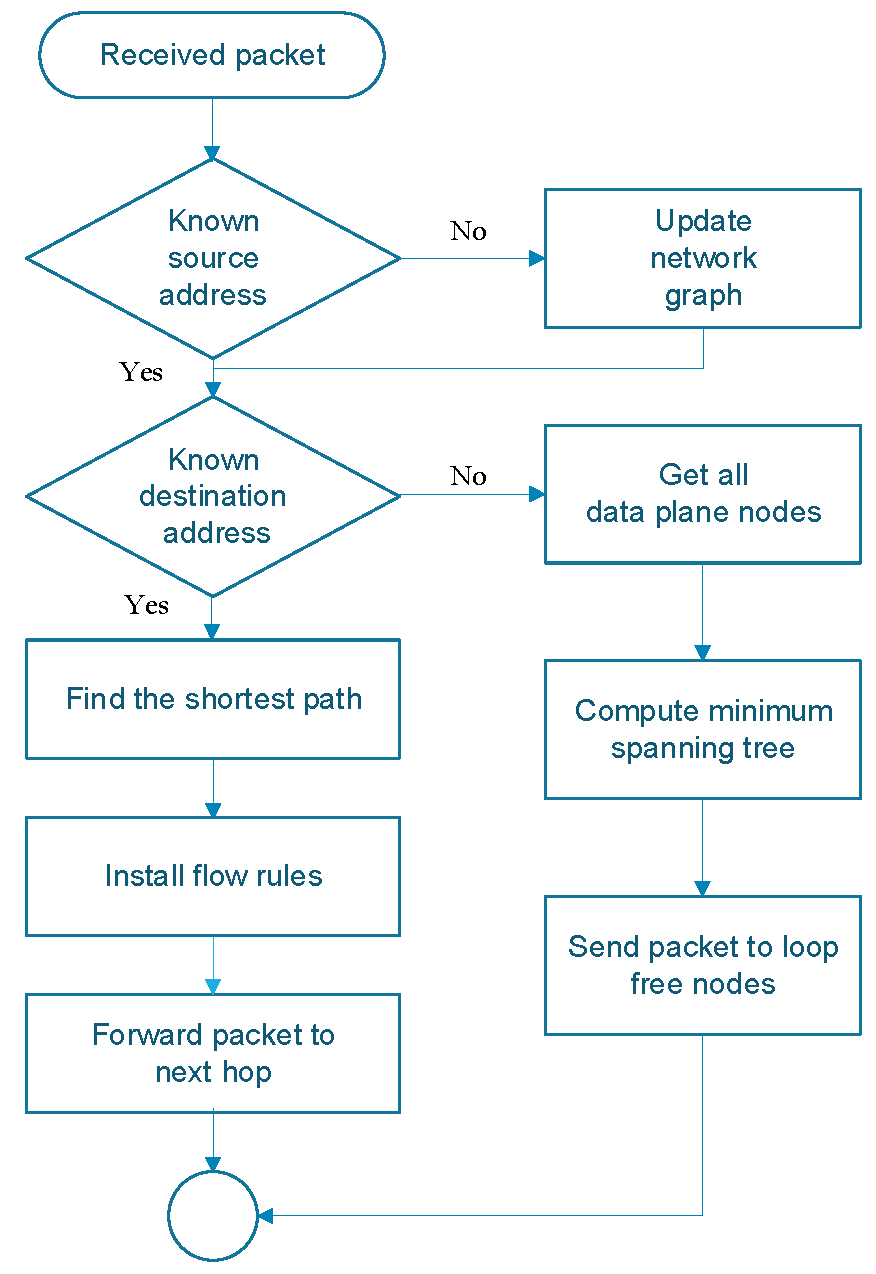}
\caption{Path selection algorithm \cite{59}}
\label{Fig:Flowchart}
\end{figure}

Battlefield communication network requires consistent adjustment of loads and routing paths while transferring video streams or performing surveillance operations using limited bandwidth of uneven availability. For example, in \cite{59}, the dynamic optimization of end-to-end paths between the source and the destination is exploited for adaptive video streaming in the battlefield network. The path selection algorithm applied adopted in \cite{59} is depicted in Fig. \ref{Fig:Flowchart}. Additionally, in \cite{60}, an adaptive link sensing approach for an aerial battlefield network is proposed that exploits back-up routing path in case of sudden network congestion. The implications of adaptive routing for mobile military devices are also discussed in \cite{61}. It aims at virtualizing the network functions at the granular level to enhance network survivability.  

Apart from them, in \cite{62}, an adaptive tactical data collection system is developed that selects the data sourcing node according to the link availability and traffic characteristics in terms of packet rate and flow distribution. When the network resources are limited, the system autonomously reduces the rate of data transmission. It also helps to reduce the amount of duplicate data and improves the accuracy of data analysis. Moreover, to balance the load among distributed controllers, a self-adaptive technique is proposed in \cite{63}. It dynamically migrates switches from one controller to another considering the geographical boundary and variation of loads. The scheme triggers based on a threshold of packet arrival rate to the controllers which can also be adjusted as per the context of the network resources. However, these existing adaptive solutions are highly suitable for the applications which have already been customized to run in SDN. For legacy applications, they provide a very narrow scope for further service enhancement.

\subsection{Network Slicing} \label{sec:sub1end}
Through network slicing, operators can create unique but logical partitions of a physical network infrastructure and simplify their multiplexing for end-to-end communications. Network slices can be expanded across different network domains such as access, core, and transport, and can be exploited to meet diverse requirements of a particular application \cite{NetworkSlice}. It harnesses both SDN and NFV concepts to increase service flexibility within the network. Since network slices are isolated, they inherently avoid the control plane congestion of one slice to affect the other slices. Moreover, every network slice maintains a set of resource and network function management policies to address speed, capacity, connectivity, and coverage-driven issues. Unlike virtual private network (VPN), network slicing does not solely rely on tunneling. It also differs from Differentiated Services (DiffServ) as noted in Table \ref{Tab:SliceDiff}.

Different SDN-enabled frameworks harness the concept of network slicing for offering better services. For example, in \cite{64}, an end-to-end network slicing framework incorporating a virtual resource manager is proposed that places network slices over physical resources based on the data traffic pattern, user connectivity demands and channel bandwidth. The resource manager can also deal with the sudden surges in resource demand and offers scope for integrating real-time decision-making policies. In another work \cite{65}, a data-driven resource management framework for network slices is proposed. The resource cognitive engine of the framework collects the resource usage data and incorporate a machine learning technique for their uniform scheduling. Conversely, the service cognitive engine analyses the user’s requirements and interact with the global cognitive engine for improving the resource utilization and user’s quality of experience. Similarly, in \cite{66}, a machine learning-based network slicing framework is proposed that divides each logical slice into a set of virtualized sub-slices and orchestrate them with different prioritized resources as per the application requirements. The framework also engages separate sub-slices to handle spectral efficiency, low latency service delivery, and power consumption, and uses the Support Vector Machine (SVM) algorithm to extract the features of assigned applications. Nevertheless, in literature, very few research initiatives have been found that focus on augmenting network slicing with military applications. To address this gap, a set of military services including push-to-talk, cellular convergence, prioritized on-demand access, satellite backhaul for redundancy and signal jamming are identified in \cite{67} where network slicing can be easily adopted for improved performance, security and availability. However, the explicit isolation of network slices makes the coordination of security policies difficult and can lead to a breach of confidentiality in battlefield communication \cite{68}. 
\begin{table}[!t]
\centering 
\small
\caption{Differences between Network slicing and differentiated service}\label{Tab:SliceDiff} 
\begin{tabular}{|p{5.5 cm}|p{5.5 cm}|}
\hline 
Network Slicing & Differentiated Service \\\hline
Allows multiple logical networks to run on top of a shared physical network & Controls and classifies network traffic to set their flow precedence \\\hline
Simultaneously deals with the networking, computation, and storage aspects of the underlying resources & Only deals with the networking aspect of the underlying resources \\\hline
Can isolate traffic of one tenant from others and supports optimum grouping of the traffic	& Cannot discriminate the same type of traffic coming from different tenants \\\hline 
\end{tabular}  
\end{table}   
\subsection{Service Function Chaining (SFC)} \label{sec:sub2start}
\begin{figure}[!t]
\centering 
\captionsetup{justification=centering}
\includegraphics[width=\textwidth]{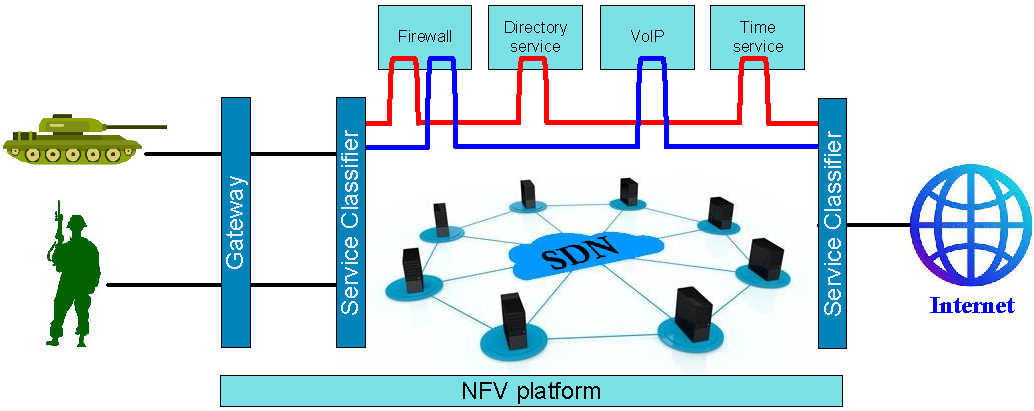}
\caption{Service function chaining in battlefield communication}
\label{Fig:SFC}
\end{figure}

Service function chaining refers to a complete suite of connected virtual network services such as firewalls, VoIP, directory service, deep packet inspection, load balancer and time service that allows traffic to use any combination of them as per the requirements in terms of security, lower latency and enhanced service quality. It also enables SDN controllers to customize a chain and apply them to different traffic flows depending on the source, destination, or type of traffic. Fig. \ref{Fig:SFC} provides an abstract representation of service function chaining for battlefield communication. 

In the literature, there has been notable initiatives that focus on improving virtual network function placement in SFC. For example, a Mixed Integer Linear Programming (MILP) model to minimize the intra-communication delay between different network function instances is proposed in \cite{70}. It meets diverse carrier-grade requirements such as latency and resource availability for an application requesting to access the service chain. In \cite{69}, another MILP model for optimizing energy consumption across multiple network domains is proposed. It considers the order of accessing the chain as a constraint and sets a domain-level function graph to orchestrate the incoming network service requests. The SDN-based resource management architecture developed in \cite{71} also aims at optimizing energy usage while placing different network functions over the computing instances and defining their routing path. As supplements, some other works are developing SFC-constrained shortest path service access mechanisms for SDN. In \cite{72}, such a mechanism is proposed that transforms the basic network graph to an SFC-constrained network graph. Moreover, it applies a pruning algorithm based on service dependency for reducing the size of newly generated network graph so that the shortest path can be calculated in timely manner. In another work \cite{73}, simple breadth-first search algorithm has been adopted to determine the shortest path. There also exists a performance evaluation framework named SFCPerf \cite{74} to check the compatibility of these approaches in real-world test bed. However, the existing solutions have significant configuration complexity that make them infeasible to deal with the instant demands of battlefield communications. 

\subsection{Unikernel Network Functions} \label{sec:sub2end}

\begin{figure}[!t]
\centering 
\captionsetup{justification=centering}
\includegraphics[width=\textwidth]{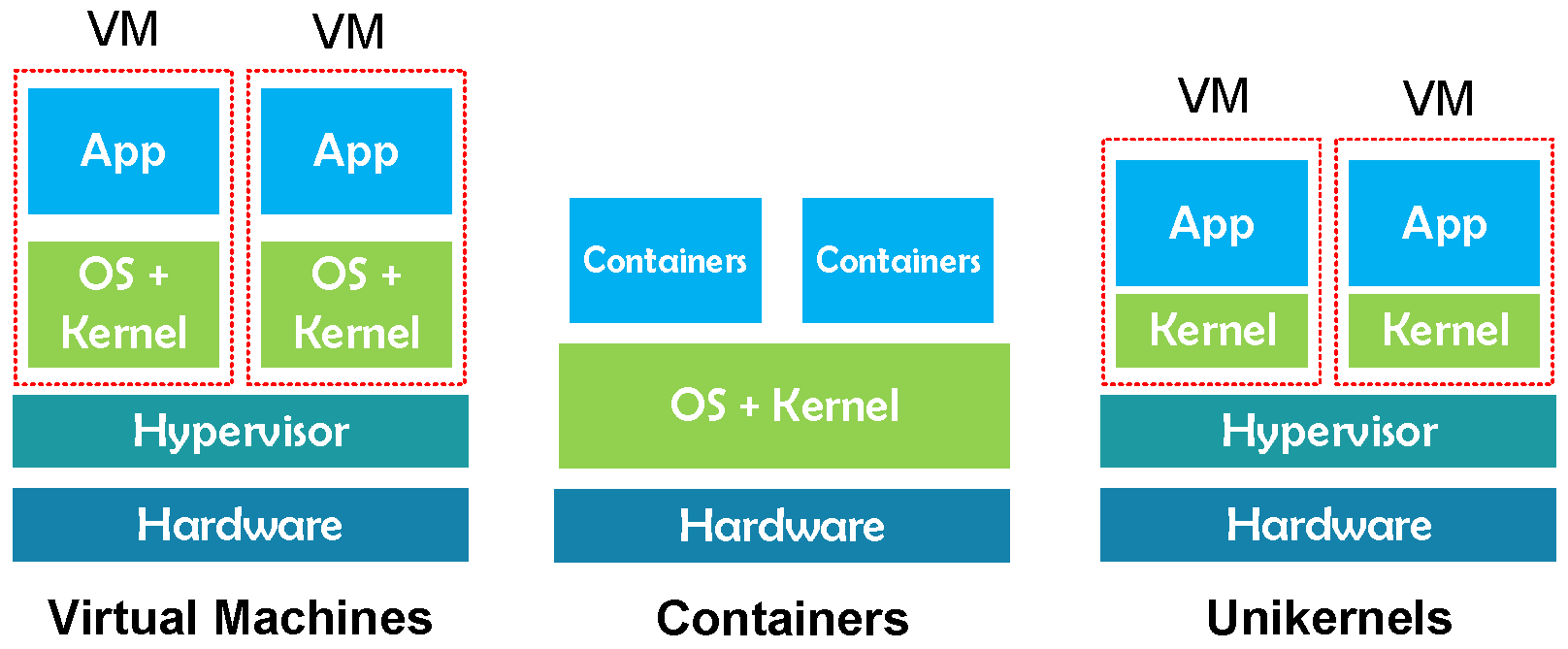}
\caption{Architecture of Virtual Machines, Containers and Unikernels}
\label{Fig:UNF}
\end{figure}

Besides virtual machines and containers, unikernels are also increasing in popularity as a virtualized software platform for implementing NFV. Unikernels refer to single-address-space machine images that can run on standard hypervisors by exploiting only kernel space libraries. The structure of unikernels is considerably lightweight compared to that of VMs, and containers, thus they can boot faster. Moreover, a unikernel can execute a single process at a time, which consequently results in less management and processing overhead. Fig. \ref{Fig:UNF} illustrates the architectural differences between VMs, containers and unikernels. Because of the low memory footprint and initiation time, unikernels are considered more well-suited for network function virtualization than VMs and containers, especially when they are used to complement any SDN-enabled system.

The concept of unikernel is relatively new and its standards are still evolving. In \cite{75}, an SDN-enabled framework is developed that can create unikernels dynamically. It enhances system reliability with respect to anomaly or security attacks and helps in recovering the system functionalities within minimal time. Similarly, in \cite{76}, the Topology and Orchestration Specification for Cloud Applications (TOSCA)-language has been extended to support the creation and orchestration of unikernels with security constraints. It also enables the unikernels to offer on-demand network services to the users. In another work \cite{77}, the initiation time of different unikernel-based network services is optimized by consistently modifying their schedulers according to the service requirements. Although unikernels outperform VMs and containers in various aspects, the packet loss rate with unikernels is higher than others. This limitation of unikernels can affect any battlefield communication requiring high throughput.  
\section{Traffic Management}
Quality-of-Service (QoS) and Quality-of-Experience (QoE) related traffic management has been studied for many years, and a significant amount of research has been devoted to understanding, measuring, and modelling QoS/QoE for a variety of network services \cite{QoE}. Considering different network segments, disparate application needs, and multiple transmission bearers involved in the end-to-end service delivery chain, it is challenging to identify the root causes of service quality impairments. It also increases the complexities in finding effective solutions for meeting the end users' requirements and expectations in terms of service quality. We briefly survey state-of-the-art findings and present emerging concepts and challenges related to managing service quality for networked services, especially in the context of the move towards softwarised networks, the exploitation of big data analytics and machine learning, and the steady rise of new application services (e.g. multimedia, augmented and virtual reality). We address the implications of such paradigm shifts in terms of new approaches in QoS modelling and the need for novel monitoring and management infrastructures.

\begin{figure}[!t]
\centering 
\captionsetup{justification=centering}
\includegraphics[width=\textwidth]{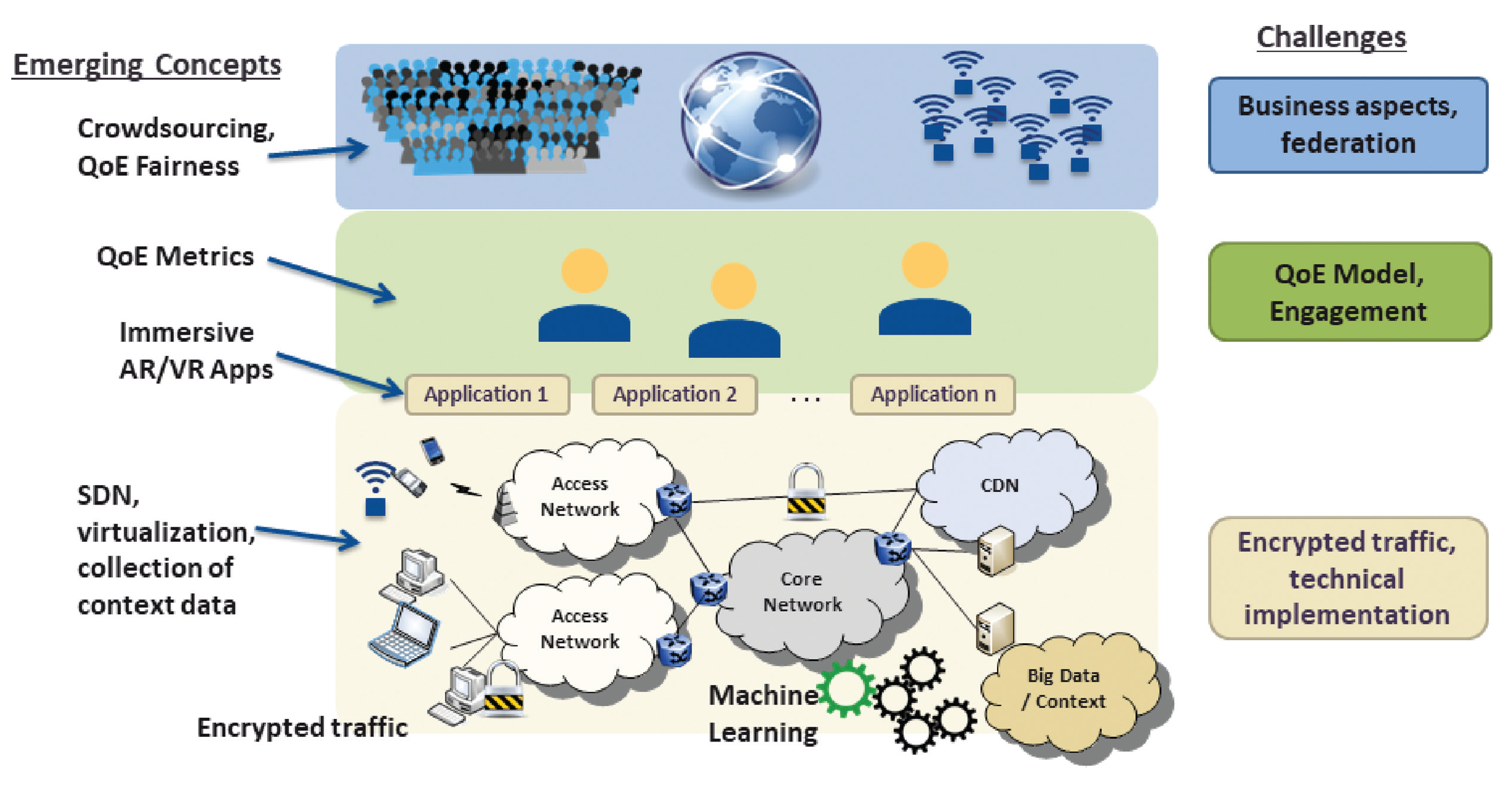}
\caption{Emerging concepts and challenges in QoS management \cite{78}}
\label{Fig:One}
\end{figure}

Traditionally, QoS-driven application management has primarily addressed control and adaptation on the end-user and application host/cloud level, often studied from an application provider perspective in the context of optimizing the quality of Over-The-Top (OTT) applications and services. As an example, applications such as HTTP-based adaptive video streaming dynamically adapt to varying network conditions to maintain a high level of QoS. Such a mechanism represents an application control loop that is often independent of network management mechanisms. On the other hand, network providers generally rely on performance and traffic monitoring solutions deployed within their access/core network to obtain insight into impairments perceived by end users. QoS-driven network management mechanisms have thus focused on the network provider point of view and considered control mechanisms, such as optimized network resource allocation, admission control, QoS-driven routing, and so on. Such control thus aims to facilitate efficient network operations and maintain high QoS, without directly managing the applications.

SDN serves as a technology for decoupling hardware resources from software and functionality, enabling programmability of the networking infrastructure. The programmable and flexible resource allocation, coupled with softwarisation, enable the network and application to engage in a "conversation" using software APIs. While this explicit negotiation approach offers clear opportunities, there are many challenges that need to be addressed (as shown in Fig. \ref{Fig:One}), including encryption of traffic, virtualization of resources, contextualization of application data, measurement of service quality, fairness, business arrangements, and federation across networks. In what follows we briefly review the evolution of QoS traffic management and recent directions enabled by SDN.

\subsection{Service Level Agreement (SLA)-aware Traffic Management}

\begin{figure}[!t]
\centering 
\captionsetup{justification=centering}
\includegraphics[width=0.8\textwidth]{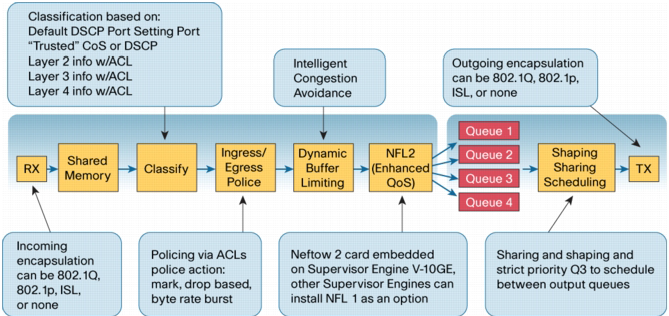}
\caption{Mechanisms for implementing SLA-based QoS \cite{79}}
\label{Fig:Two}
\end{figure}

The notion of using service level agreements (SLAs) for QoS dates back to the IETF IntServ and DiffServ frameworks \cite{79}, whereby the application specifies its requirements in the form of a FlowSpec, which includes both its traffic profile (rate and burstiness) and requirements profile (in terms of guaranteed bandwidth and latency) – once accepted by the network (via some form of admission control), this forms an agreement (SLA) that then needs to be respected by both parties. The realization of this framework (as shown in Fig. \ref{Fig:Two}) requires admission control (often via a bandwidth broker), traffic classification (using packet header fields), packet marking (typically as a DiffServ Code Point or DSCP), traffic policing (via a token bucket), and priority or weighted fair scheduling to ensure network resources are shared in order to meet the pre-negotiated SLAs.

While conceptually elegant, the major challenges with this approach relate to the large amount of state information along with the complex policing/scheduling mechanisms needed for managing the per-flow SLA, as well as limitations in being able to map application-level QoE to network-level QoS parameters – these aspects are explored in depth in \cite{80}, which also develops a new method called SFQP (SLA-aware Fine-grained QoS Provisioning) to perform the mapping and bandwidth enforcement using SDN principles. Other works including \cite{81} have also explored the application of QoS methods enabled by SDN protocols (OpenFlow in particular) to support the classification, prioritization, and shaping of application flows with a view towards enabling dynamic QoS control.

In networks where the applications are not enabled with capabilities to explicitly negotiate SLAs, the application behavior as well as requirements may need to be inferred. The work in \cite{82} develops an application-aware traffic engineering system that cooperates with deep packet inspection (DPI) services to apply SDN based prioritization and route selection to application flows. A specific application of this concept to VoIP and M2M communication in developed and demonstrated in \cite{83}, whereby it is shown that SDN can be used to proactively manage UDP/RTP media streams to enhance their service quality.

\subsection{Intent-based Traffic Management}

Intent-based networking (IBN) is a relatively new concept in SDN for managing a network, end-to-end, through the use of DevOps and high-level "intents". The term IBN was first coined by Gartner in 2017, though components of intent-based networking began well before and continue to be developed by networking enterprises. Traditional networking relied on command line interface (CLI) to manually set up policies for all vendors' networking devices individually. The intent-based networking approach changes this to operate it as a Network-as-a-Service (NaaS), meaning it is end-to-end networking that seamlessly manages all devices on one interface. While similar to the principles of SDN, IBN differs by integrating DevOps into the process. This makes networking management a lifecycle process that, according to Cisco, "bridges the gap between business and IT."

\begin{figure}[!t]
\centering 
\captionsetup{justification=centering}
\includegraphics[width=0.8\textwidth]{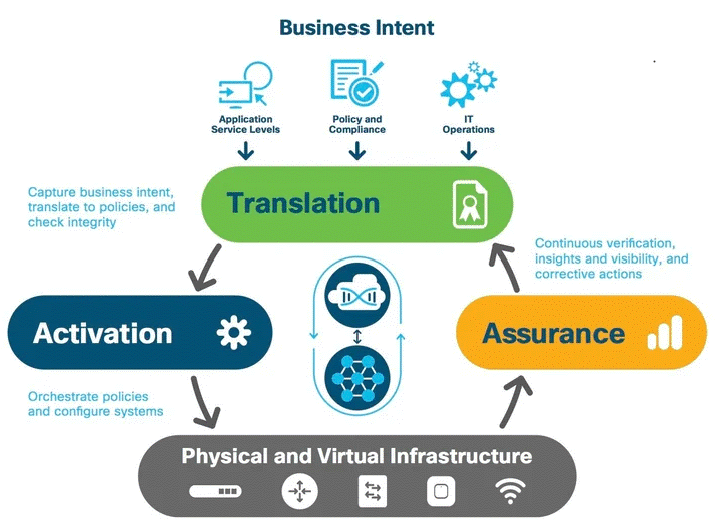}
\caption{Elements of Intent-Based Networking (source: Cisco)}
\label{Fig:Three}
\end{figure}

As a simple example of IBN, consider an Intent whereby the network operator wants to ensure that the command and control (C\&C) communications in the region receive uninterrupted service levels during combat (as shown in Fig. \ref{Fig:Three}). The Translation of this would build a policy which guarantees that C\&C users and applications are placed on a secure segment that receives the highest priority service. The Activation of this intent may apply priority-service levels between all users and applications on the C\&C bearer segment across all network elements. The Assurance module will use telemetry to monitor and analyze the network against this desired outcome, to remediate, optimize, and correct as appropriate. In order for intent-based networking to achieve its full potential, these functions are applied across all networking domains and build on a programmable network infrastructure.

Intent-based networking is being incorporated into many of the emerging SDN platforms. Both the Open Networking Operating System (ONOS) and the OpenDayLight (ODL) SDN controllers incorporate "intents". An example framework for intents is specified by Group Based policy (GBP), which has the concept of end-point groups (EPGs) so that policies can be applied to groups of entities based on their labels, and the policies themselves are contracts with "qualities" and "clauses".  

\begin{figure}[!t]
\centering 
\captionsetup{justification=centering}
\includegraphics[width=0.6\textwidth]{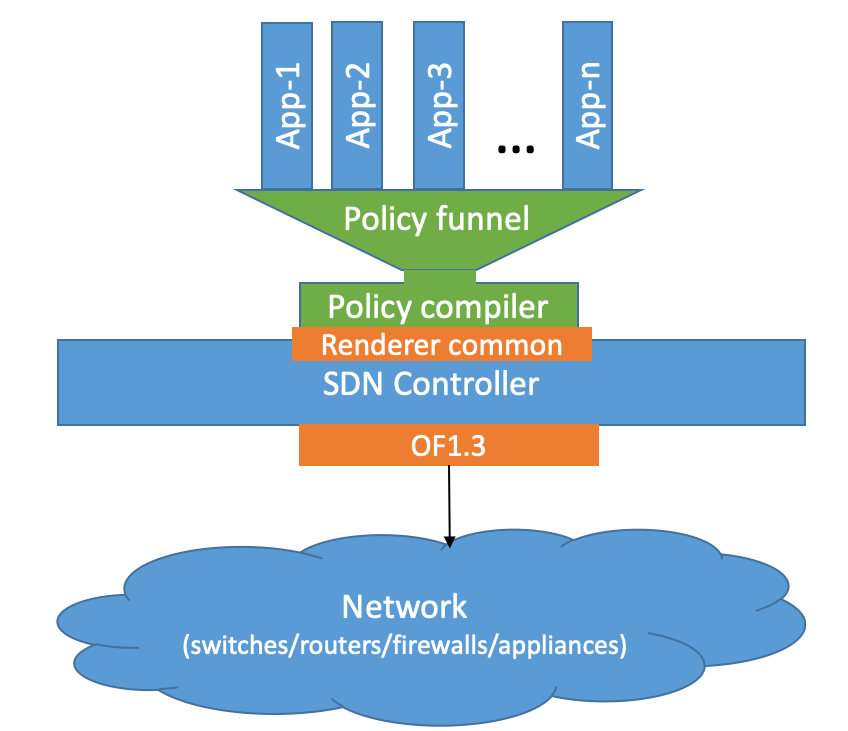}
\caption{Intent-based Networking supporting multiple applications}
\label{Fig:Four}
\end{figure}

One of the significant benefits of using high-level intents rather than low-level network configuration is that human errors are reduced. The high-level intents are automatically "compiled" by a policy compiler that translates the intents into network device configuration, which is pushed down to each network element. Further, multiple applications can co-exist without conflict; as shown in Fig. \ref{Fig:Four}, application policies are taken through a policy funnel into a compiler that flags, and potentially automatically resolves, any conflicts in their policies. Apstra reports in \cite{84} that IBN can be applied in a vendor- and technology-independent way, yielding a saving of seven cents per dollar revenue.

\subsection{Context-aware Traffic Management}

Context-aware traffic management is emerging as an approach to address some of the gaps in SLA and intent-based methods. The SLA-aware method requires applications to specify their requirements, which can be very challenging especially when they are adaptive themselves. The intent-based methods also need to be aware of context, such as whether the network is operating in a friendly or hostile environment. The context-aware approach considers the "experience" of the application, couples that with the context, and takes reactive actions to rectify the problem.

\begin{figure}[!t]
\centering 
\captionsetup{justification=centering}
\includegraphics[width=0.8\textwidth]{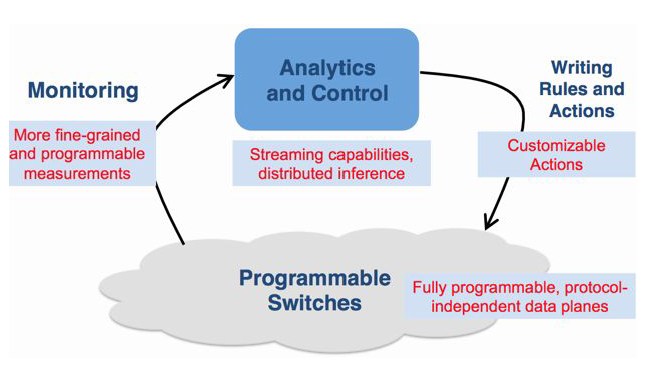}
\caption{Self-Driving Network with Monitor-Analyse-Control loop \cite{85}}
\label{Fig:Five}
\end{figure}

This thinking is leading to the concept of a "self-driving network" \cite{85} as depicted in Fig. \ref{Fig:Five}, whereby the network is continually monitored using fine-grained telemetry, the collected data is analyzed in real-time, and appropriate intervention is done via programmable network interfaces to take an appropriate control action. Research work in \cite{86} develops a framework for adjusting network behavior dynamically to adapt to application behavior and validates it via implementation on multiple SDN switches in \cite{87}. Conceptually, both Self-driving networking and Intent-based networking aims at autonomic management of the network. However, intent-based networking consistently tunes the networking environment as per the user's feedback whereas, self-driving networking monitors the differences of the current and the desired network state and tunes the networking environment accordingly.

Google has demonstrated that it is able to adapt its traffic management across data centres \cite{88}, within a data centre \cite{89} \cite{90}, and throughout its peering locations \cite{91} using dynamic application level measurements and fine-grained SDN control. Network operators are stymied in this effort due to lack of visibility into application performance, compounded by the increasing encryption of packets by application – however, new methods are being developed by a research team that use machine learning-based methods to identify applications \cite{92} and infer experience \cite{93}, and further take corrective action reactively when application experience shows symptoms of degradation \cite{94}. Moreover, QoS-aware traffic management is progressing towards this virtuous cycle of a self-aware network that constantly monitors application experience, makes inferences based on operator-supplied intents combines with contextual information, and then enforces control into the programmable network substrate in an automated manner. 

\section{Policy Evaluation}\label{Sec:LitEnd}

There are different ways to evaluate the efficiency of SDN-based policies such as empirical, emulation and simulation.  Empirical analysis refers to an evidence-based approach that relies on real-world implementation and results. From the perspective of SDN, empirical analysis is an essential. However, since an SDN environment incorporates numerous entities interacting with each other across control, data and application plane, the real-world implementation of SDN for research is costly. Moreover, modification of any entity in real-world implementation is tedious. In this case, emulation or simulation can be adopted for approximate imitation of SDN-based operations. Emulation duplicates the behavior of the real system whereas simulation mimics the behavior but does not offer the exact matching. In the following subsections, the recent practices on empirical, emulation and simulation-based analysis of SDN operations are discussed.  

\subsection{Empirical}

There has been a notable initiative in SDN that focuses on empirical evaluation of policies. For example, in \cite{95}, a small-scale software defined cloud datacenter named CLOUDS-Pi is developed. To enable Raspberry Pi devices as network switches, CLOUDS-Pi augments Open vSwitch (OVS) with each of them and uses OpenDaylight (ODL) as the SDN controllers. Through use case study, it has also been illustrated that CLOUDS-Pi is capable of evaluating the performances of any SDN-based virtual machine management and flow scheduling policies. In another work \cite{96}, the performance of seven SDN switches (as noted in Table \ref{Tab:Spec}) are benchmarked in terms of throughput, priority queuing, flow tables and packet buffers. It has also been observed that the processing time of the switches is predictable and is aligned with the line rate. Moreover, in \cite{97}, a publicly available bug repository for OpenDaylight SDN controller is mined to localize the most problematic software components and model the stochastic behavior of bug manifestation. Later, the information is applied to improve the dependability of different components such as core controller functions, embedded applications, plug-ins, and drivers in the control plane. Furthermore, the effect of strong and eventual consistency constraints on scalability and correctness of control plane is investigated in \cite{98}. It has also evaluated an adaptive consistency model that improves the request handling throughput and response time of controllers. However, because of large-scale and sophisticated deployment of SDN components, the arrangement of empirical analysis in battlefield communication is often regarded as infeasible.

\begin{table}[!t]
\centering 
\small
\caption{Specifications of the investigated switches \cite{96}}\label{Tab:Spec} 
\begin{tabular}{|p{1.9 cm}|p{2.6 cm}|p{3 cm}|p{3.6 cm}|}
\hline 
Switch	& ASIC & CPU & Firmware (release date) \\\hline
HP E3800 &	HPE ProVision & Freescale P2020 &	KA.16.04.0016 (2018-06-22) \\\hline
HP 2920	& HPE ProVision &	Tri Core ARM1176 &	WB.16.08.0001 (2018-11-28) \\\hline
Dell S3048-ON &	Broadcom StrataXGS & undisclosed & DellOS 9.14 (2018-07-13) \\\hline
Dell S4048-ON &	undisclosed & undisclosed & DellOS 9.14 (2018-07-13) \\\hline
Pica8 P3290	& Broadcom Firebolt 3 & Freescale MPC8541CDS &	PicOS 2.10.2 (2018-01-19) \\\hline
Pica8 P3297	& Broadcom Triumph 2 & Freescale P2020	& PicOS 2.11.19 (2019-02-27) \\\hline
NEC PF5240 & undisclosed & undisclosed & OS-F3PA6.0.0.0(2014-06) \\\hline
\end{tabular}   
\end{table}   

\subsection{Emulation}

\begin{figure}[!t]
\centering 
\captionsetup{justification=centering}
\includegraphics[width=\textwidth]{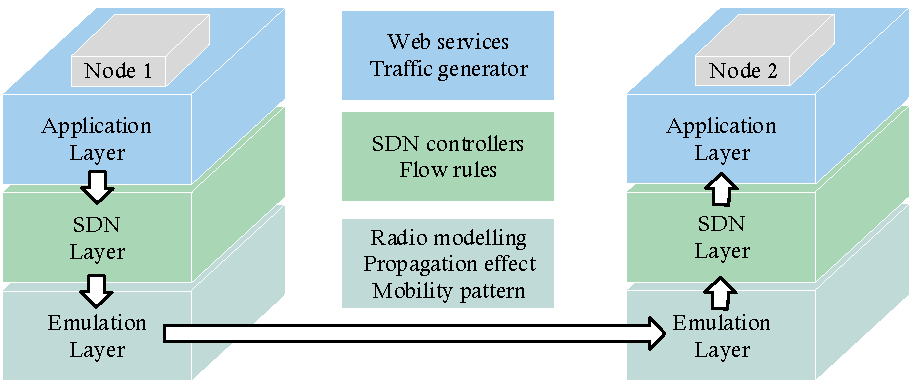}
\caption{Node to node communication within an emulated tactical SDN \cite{102}}
\label{Fig:Emu}
\end{figure}

As noted, military tactical networks require to support mission-critical operations in the austere environment by going beyond the mobility, intermittent link state, and variable bandwidth-related issues. In real-world SDN environments, the manifestation of such dynamic configurations for research purpose is extremely challenging. Therefore, it is widely adopted to imitate military tactical networks using different emulation tools such as Emane \cite{99}, Mininet \cite{100} and Core \cite{101}. An emulator simultaneously captures the characteristics of tactical communications and integrates SDN methodologies to assess different control and management policies over an imitated military tactical network \cite{102}. Fig. \ref{Fig:Emu} depicts how emulators can be augmented in node-to-node communications. 

Among the SDN emulators, Mininet is the most popular. In the literature, Mininet has been adopted to evaluate policies for deploying SDN controllers \cite{53}, enhancing controller's adaptivity \cite{54}, automating distributed firewalls \cite{57}, managing data flow \cite{58}, augmenting Named Data Networking (NDN) \cite{103} and creating integrated SDN environments \cite{104} in tactical networks. Mininet is lightweight, boots faster and offers higher scalability. However, it is difficult to employ Mininet for dealing with non-Linux-compatible OpenFlow switches or applications.

Extendable Mobile Ad-hoc Network Emulator (Emane) is another celebrated emulator for tactical networks which has been used in \cite{105}, \cite{106}, \cite{107} and \cite{108} to evaluate various policies for group-based communications, latency-aware queuing control, situation-aware publish subscribe model and mission-centric content sharing respectively. Emane incorporates more detailed radio models that simplify the emulation of MANET, although it lacks an accurate interference model based on Signal-to-Interference-and-Noise-Ratio (SINR) and extensive libraries for imitating complex scenarios in SDN environments.  

There exists another emulator named Common Open Research Emulator (CORE) that has been used in evaluating policies for delay tolerant routing \cite{109}, data and control plane security management \cite{110}, and disruption-tolerant networking \cite{111}. CORE offers highly customizable programming interfaces that simplifies its augmentation with other emulators including Emane. However, it lacks facilities for distributed emulation. Apart from Emane, Mininet and Core, there exists another emulator named Containernet which has been used in \cite{112} for hybrid service function chaining. 

\subsection{Simulation}

The existing emulators for SDN mainly focus on network resources management and provide a very limited scope to apply application and computing resource-level management techniques such as service placement and resource consolidation. To address this issue, different simulators such as OPNET, NetSim and CloudSim-SDN are used in SDN-based policy evaluation. Among them, OPNET is used in \cite{113} for simulating data distribution in a tactical network. In \cite{114} and \cite{115}, OPNET is also adopted to evaluate a cooperative trust scheme and QoS-aware routing policy for military communications respectively. Although OPNET provides a set of extensive libraries for detailed networking models, it lacks support for customization.

\begin{figure}[!t]
\centering 
\captionsetup{justification=centering}
\includegraphics[width=\textwidth]{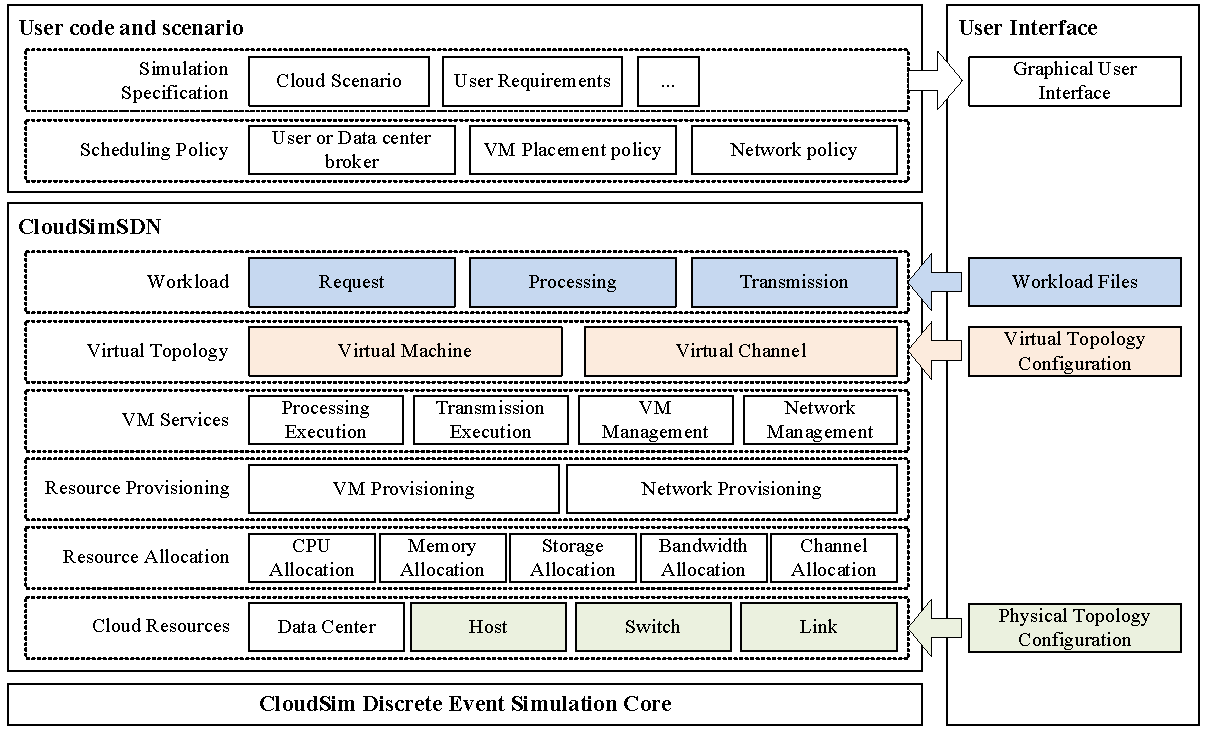}
\caption{Overview of CloudSim-SDN}
\label{Fig:CSim}
\end{figure}

Like OPNET, NetSim is used in simulating different network and application management scenarios. For example, in \cite{116}, a hybrid routing policy for MANET and in \cite{117}, an intrusion detection framework for military communication is evaluated through NetSim. One of the main advantages of NetSim is that it can simulate the functions of a wide range of networking devices. On the other hand, the operations of NetSim are handled by a single event queue that often resists the modeling of complex scenarios. Similar to NetSim, CloudSim-SDN is another discrete event simulator \cite{118}. It has been developed by the Cloud Computing and Distributed Systems (CLOUDS) Laboratory, University of Melbourne. As noted in Fig. \ref{Fig:CSim}, CloudSim-SDN runs on top the basic CloudSim simulator \cite{119} that allows users to model both physical and virtual topology, and application scenarios \cite{MCCDef}. Using this feature of CloudSim, different simulators for other computing paradigms for example iFogSim \cite{iFogSimTut} and MR-CloudSim \cite{MRSIM} have also been developed. However, using CloudSim-SDN, a user can either utilize built-in resource management and scheduling policies or can develop their own by extending the abstract interfaces. As a means of policy evaluator, CloudSim-SDN has been used in \cite{120} that focuses on latency-aware network function provisioning. It has also been adopted for simulating elastic service function chaining \cite{121} and energy-efficient network optimization \cite{122} policies. However, the current version of CloudSim-SDN lacks supports for handling the dynamics of tactical network but there is always a potential scope to augment them in CloudSim-SDN.

\section{Gap Analysis and Future Directions}\label{Sec:Future}
The lessons learned and the gaps identified from the literature study can be summarized as follows:

\begin{enumerate}

\item In battlefield communication or tactical networks, MANET is highly adopted because of its flexibility, ease of mobility and lower capital or operational expenses. However, the convergence of multi-bearer networking, MANET and SDN, specially for military operations, has been barely explored in the literature.

\item The device-level interactions and connectivity at the data plane of SDN-enabled tactical networks is unpredictable and unreliable. Military devices also have limited energy supply to operate \cite{ICSOC}. In such cases, dynamic network partitioning and fault tolerance techniques can be useful in supporting the vulnerable military devices losing connections with the controllers. However, these aspects have been addressed by very few research initiatives in the literature. Additionally, there is a significant lack for emulation and simulation tools to imitate such scenarios specifically for military use cases.  

\item Inherently, the controller is a single point of failure for the entire SDN architecture. To deal with this issue, the concept of multi controllers in SDN has been developed. However, the existing East–West communication mechanisms between the controllers still follow the traditional centralized architecture and cannot ensure robust spanning of network through flat controller orientation. Moreover, the Northbound and Southbound interfaces for multi-controller SDN architectures are currently poorly defined and hinder the real-time integration of the management systems and the peer-level networks. These constraints affect the multi-domain communications, slice management and intent-based networking in tactical environments, especially when one ground military device sends information to an aerial or submerged military device. To address such scenarios, efficient multi-controller orchestration policies must be developed according to the requirements of battlefield communications. 

\item The subordinates of a tactical network are hierarchically arranged. At lower levels, line-of-sight connectivity is operated by distributed wireless mesh (MANETs). Multiple MANETs can also coexist at this level with thin Inter-MANET connectivity. At the mid hierarchical levels, satellite communication techniques are harnessed, whereas a mix of terrestrial wireless, SATCOM and wireline connectivity is exploited at the higher levels of the tactical network. Most of the mechanisms are well suited for legacy network and provide a narrow scope to integrate SDN functionalities. In the literature, there is also a significant lack in building interoperability among MANET, terrestrial wireless, and satellite communication techniques, especially through SDN middleware. 

\item A wide range of traffic from real-time (e.g. situation and location-aware dissemination) to elastic (e.g. audio or video files) is generated during battlefield communication. This traffic can be both unidirectional and bidirectional between mobile (e.g.  tanks and submarine) and fixed entities (e.g. ground stations). To meet QoS requirements under such diverse circumstances, different sophisticated and adaptive traffic management schemes are required for tactical networks. These schemes should also support the analysis of ingress/egress packets and the appropriate selection of differentiated services and network slices. On the other hand, the efficiency of these schemes is highly subjected to the QoS requirements of the SDN-enabled applications and the security concerns of the battlefield communications. However, in the literature, the quantification of QoS parameters and security classifications with respect to tactical networks and the management of traffic in accordance are narrowly explored.

\item As noted, intent-based networking allows users and operators to define their service expectations from the network and simultaneously creates the desired networking state for meeting those expectations. The ultimate goal of intent-based networking is to reduce the complexities of enforcing various network management policies. However, the augmentation of intent-based networking with traditional SDN architecture requires a comprehensive synthesis of artificial intelligence (AI), network automation and machine learning (ML). On the other hand, autonomic network management depends on four different aspects: (\textit{i}) Self-configuration: configures the network components (e.g. nodes and bandwidth), (\textit{ii}) Self-healing: treats the faults and adapts with the dynamics, (\textit{iii}) Self-optimization: enhances performance of the networking components, (\textit{iv}) Self-protection: protects from the security attack. Nevertheless, in the literature, these essential aspects of intent-based networking have not been fully investigated with respect to tactical networks.

\end{enumerate}

\section{Summary}\label{Sec:Done}
The concept of SDN is gradually attracting attention in military use cases. However, the adoption of SDN in tactical network is subjected to diverse challenges with respect to interoperability, distributed application, unpredictable service demand, security constraints and edge computation. Although there exist a notable number of works on the literature aiming at addressing these challenges, they have certain limitations and compatibility issues with existing tactical communication standards such as MBN and MANET. In this work, we reviewed such research initiatives that primarily focus on the SDN-based network orchestration problem in the tactical environments. We proposed a taxonomy to categorize the existing solutions systematically and determined the research gaps for further improvement in this domain. 
\section*{Acknowledgement}
This work was supported by the Next Generation Technologies Fund, managed by the Defence Science and Technology Group, the Department of Defence, Australian Government. The authors would also like to thank Shashikant Ilager and Muhammed Tawfiqul Islam for discussions and comments on improving the paper.
\bibliographystyle{splncs}
\bibliography{references}

\end{document}